\documentclass[english]{IEEEtran}
\usepackage[T1]{fontenc}
\usepackage[latin9]{inputenc}
\usepackage{color}
\usepackage{float}
\usepackage{amsmath}
\usepackage{amssymb}
\usepackage{graphicx}

\makeatletter

\providecommand{\tabularnewline}{\\}
\floatstyle{ruled}
\newfloat{algorithm}{tbp}{loa}
\providecommand{\algorithmname}{Algorithm}
\floatname{algorithm}{\protect\algorithmname}

\usepackage{cite}
\usepackage{times}
\usepackage{URL}

\usepackage{algorithmic}

\usepackage{subfig}

\ifdefined\showcaptionsetup
 \PassOptionsToPackage{caption=false}{subfig}
\fi
\usepackage{subfig}
\makeatother

\usepackage{babel}
\begin{document}
\title{A Group-Wise Narrow Beam Design for Uplink Channel Estimation in Hybrid
Beamforming Systems}
\author{{\normalsize Yufan Zhou, Yongbo Xiao, and An Liu, }~\IEEEmembership{Senior Member,~IEEE}{\normalsize\thanks{Yufan Zhou, Yongbo Xiao, and An Liu are with the College of Information
Science and Electronic Engineering, Zhejiang University, Hangzhou
310027, China (email: yufanzhou@zju.edu.cn, 22231161@zju.edu.cn, anliu@zju.edu.cn).}}}
\maketitle
\begin{abstract}
In this paper, we consider uplink channel estimation for massive multi-input
multi-output (MIMO) systems with partially connected hybrid beamforming
(PC-HBF) structures. Existing beam design and channel estimation schemes
are usually based on ideal assumptions and require transmitting pilots
across multiple timeslots, making them unsuitable for practical PC-HBF
systems. To overcome these drawbacks, we propose a novel beam design
and a corresponding channel estimation algorithm to achieve accurate
and real-time uplink channel estimation. Firstly, we introduce a group-wise
narrow beam design in the vertical dimension to suppress interference
from undesired angular components and improve vertical angle estimation
accuracy, which divides the columns of the uniform planar array (UPA)
into groups and the vertical angle interval into sub-intervals. In
this way, each group is assigned with a narrow beam to cover one vertical
angle sub-interval, and the set of narrow beams is designed based
on the filter design theory. Secondly, we optimize the antenna grouping
pattern using the Estimation of Distribution Algorithm (EDA), balancing
interference suppression and resolution capability in the horizontal
dimension, leading to a better horizontal angle estimation performance.
Finally, we design a low-complexity group-wise subspace constrained
variational Bayesian inference (GW-SC-VBI) algorithm to fully take
advantage of the proposed beam design to achieve both low-complexity
and high-accurate channel estimation. Simulation results demonstrate
that the proposed scheme achieves notable performance gains over baseline
methods.
\end{abstract}


\section{Introduction}

Massive MIMO has emerged as the core physical layer technology in
5G New Radio (NR) \cite{lu2014overview}. In addition, the Hybrid
Beamforming (HBF) structure was introduced in \cite{ven_2010_HBF}
to balance performance and cost for massive MIMO by combining analog
and digital beamforming techniques. Specifically, analog beamforming
uses low-cost analog phase shifters to reduce the number of required
RF chains, while digital beamforming performs low-dimensional digital
signal processing at the baseband. Generally, the HBF structure can
be categorized into two types. The first type is the fully-connected
structure, where each RF chain is connected to all antennas through
phase shifters, enabling maximum beamforming gain \cite{ven_2010_HBF}.
In contrast, the Partially Connected Hybrid Beamforming (PC-HBF) structure
proposed in \cite{song_2019_full_with_part} further reduces costs
and complexity by connecting each RF chain to only a subset of antennas.
Due to its efficiency, the PC-HBF structure is more commonly adopted
in practical systems.

Despite many advantages of PC-HBF structures, acquiring accurate and
real-time channel state information (CSI) at the BS becomes significantly
more challenging. Traditional methods such as least square (LS) approach
\cite{SRS_based_CE} and minimum mean square error (MMSE) method \cite{singal_processing_book,rottenberg_2020_FDD}
perform poorly under HBF structures, since the dimension of the desired
CSI is larger than that of the received signal.

Numerous studies have attempted to achieve high precision channel
estimation for massive MIMO systems with PC-HBF structures by obtaining
accurate angle domain estimation. In \cite{fan_2018_angle_HBF}, the
authors designed different analog beam matrices for each timeslot
and combined the received signals and analog beam matrices across
all timeslots to obtain an equivalent observation and a full rank
equivalent analog beam matrix, enabling direct application of the
traditional 2D-DFT algorithm to estimate the angles of different paths.
In \cite{2020_tsp_CBS}, the authors introduced the convolutional
beamspace (CBS) algorithm, which partitions the angular interval into
multiple sub-intervals via bandpass filters. For each sub-interval,
uniform downsampling is performed after filtering to achieve dimension
reduction while preserving the Vandermonde structure, enabling the
direct application of high-resolution subspace methods such as MUSIC
or root-MUSIC without additional processing. In \cite{2024_tsp_CBS_for_HBF},
the authors extended the CBS algorithm to channel estimation under
FC-HBF structure, utilizing it to design the analog and digital beam
matrix. Nevertheless, the CBS algorithm also requires multiple time
slots to obtain sufficient observations to recover the full-dimensional
channel and it cannot be directly applied to the more challenging
PC-HBF structure.

Another line of works have formulated the channel estimation problem
as a compressive sensing (CS) problem, since the number of scatterers
in the environment is limited and the angular domain channel exhibits
inherent sparsity \cite{liuan_2017_tsp,dai_2018_CS}. In \cite{HBF_est_journal,HBF_est_PIMRC},
the authors conducted channel estimation in multiple stages. Specifically,
they divided the angle of arrival (AoA) and the angle of departure
(AOD) ranges into multiple sub-ranges. At each stage, a beam is designed
based on feedback from the previous stage, and the angle estimation
interval is iteratively narrowed using a modified matching pursuit
CS algorithm. In a recent work \cite{SC_VBI}, the authors proposed
a low-complexity subspace-constrained variational Bayesian inference
(SC-VBI) algorithm to obtain sparse angular-domain channels, enabling
real-time channel estimation. Both the analog and digital beam matrices
are randomly designed to satisfy the restricted isometry property
(RIP) with high probability.

However, the existing beam design and channel estimation algorithms
are generally based on ideal assumptions, leading to poor performance
in practical systems, as elaborated below. Firstly, in practice, each
user is only allocated with a single pilot symbol in one time slot
in order to support uplink channel estimation for more users \cite{wan_2023_channel_tracking,5G_NR}.
And due to the system imperfections, phase noise inevitably occurs
in the received SRSs \cite{wan_2023_channel_tracking}, causing random
phase shifts and disrupting the time correlation of channels. As a
result, channels across different timeslots cannot be assumed invariant,
making it difficult to apply the channel estimation algorithm in \cite{fan_2018_angle_HBF},
\cite{2024_tsp_CBS_for_HBF} or similar approaches \cite{xiao_2019_CE_for_HBF}.
Secondly, performing channel estimation across multiple timeslots
slows down CSI acquisition and reduces the real-time performance of
the system. Moreover, practical massive MIMO systems typically employ
uniform planar arrays (UPAs) instead of uniform linear arrays (ULAs)
to support a larger number of antennas, and utilize partially connected
structures rather than fully connected structures to reduce hardware
costs. Consequently, many existing channel estimation algorithms,
such as those in \cite{fan_2018_angle_HBF,2024_tsp_CBS_for_HBF},
cannot be directly applied, since only a subset of elements in the
analog beam matrix are non-zero under partially connected structures,
leading to reduced degrees of freedom in beam design. Although the
SC-VBI algorithm \cite{SC_VBI} with random beams enables real-time
channel estimation, the random beam design is not sufficiently optimized
and cannot fully exploit array gains, which limits its performance
particularly in low-SNR regimes.

Considering the limitations of existing methods, we address the uplink
channel estimation problem for practical PC-HBF systems in this paper,
where the BS is equipped with a uniform planar array (UPA) lying in
the $yz$-plane and each RF chain is connected to a vertical subarray
(i.e., the received signals suffer from vertical compression), as
illustrated in Fig. \ref{fig:HBF_structure}. Such a PC-HBF structure
with vertical compression is widely used in practical systems due
to the following reasons. In real-world deployments, the BS is typically
installed at a certain height, and in most cases, The distances from
the BS to the users and their corresponding scatterers are both significantly
larger than the height of the BS, as shown in Fig. \ref{fig:BS_height},
resulting in the vertical angles constrained within a certain prior
interval $\Phi$, typically ranging from $-\frac{1}{6}\pi$ to $0$.
By leveraging this property, vertical compression can achieve a higher
compression ratio, thereby improving the efficiency of the system.
Specially, we propose a novel group-wise narrow beam design and a
corresponding low-complexity channel estimation algorithm only based
on a single uplink pilot symbol. Our approach effectively exploits
the achievable array gains under PC-HBF structures, enabling robust,
accurate, and real-time channel estimation. The main contributions
of this paper are summarized as follows:
\begin{figure}[t]
\subfloat[\label{fig:HBF_structure}]{\centering{}\includegraphics[width=45mm]{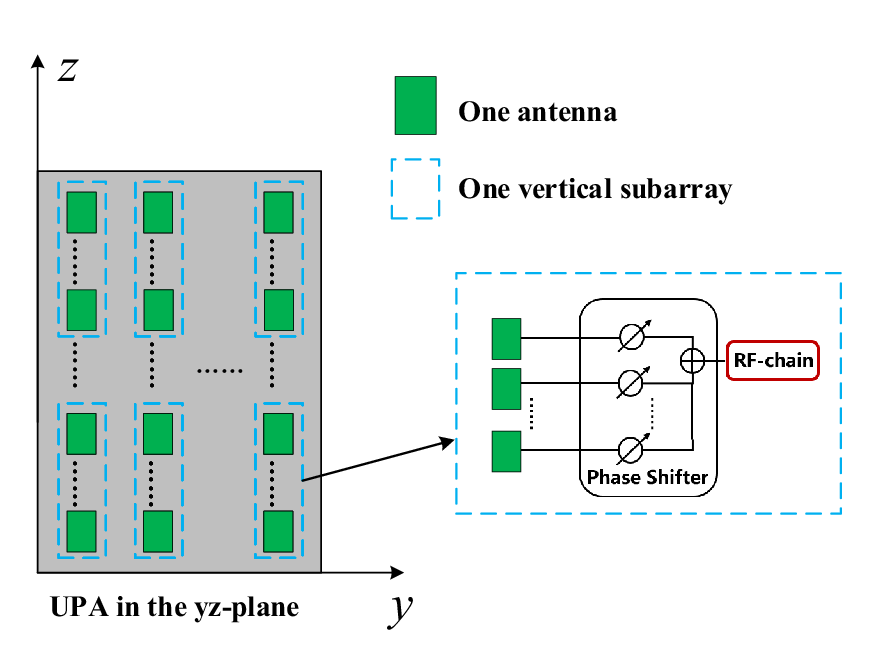}}\subfloat[\label{fig:BS_height}]{\centering{}\includegraphics[width=45mm]{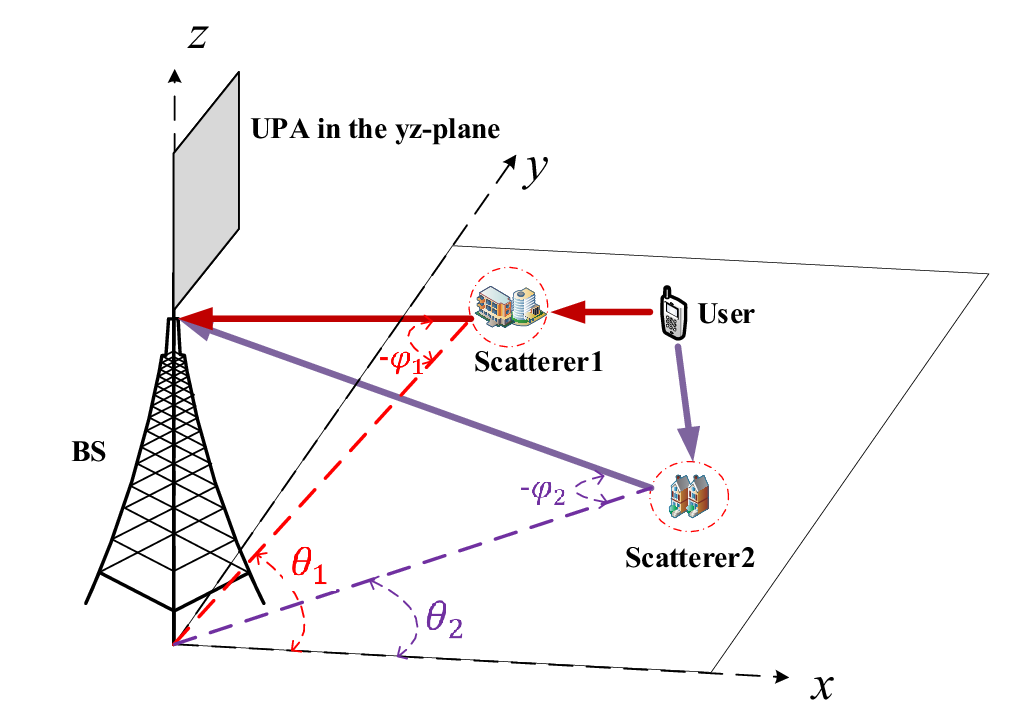}}\caption{{\small An illustration of a }real-world PC-HBF system, the negative
sign represents the vertical angle $\varphi$ is negative.}
\end{figure}

\begin{figure}
\begin{centering}
\includegraphics[width=80mm]{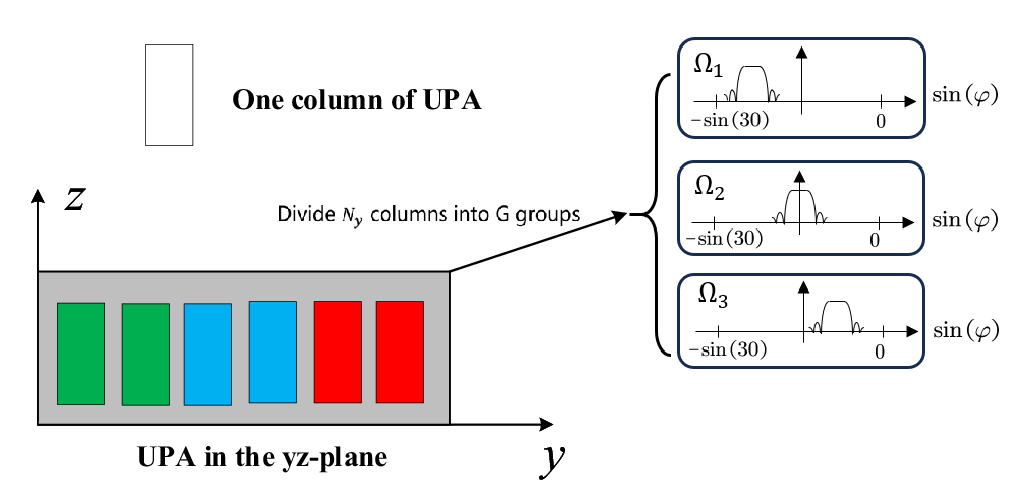}
\par\end{centering}
\caption{{\small\label{fig:group-wise}An illustration of }group-wise narrow
beam for \textbf{$N_{y}=6$} and $G=3$ with uniform grouping pattern,
and different colors correspond to different groups.}
\end{figure}

\begin{itemize}
\item \textbf{Group-wise narrow beam design in the vertical dimension:}
To tackle the limitation in vertical angle estimation performance
caused by vertical compression, we propose a novel group-wise narrow
beam design inspired by the bandpass filter banks approach adopted
in CBS algorithm \cite{2020_tsp_CBS,2024_tsp_CBS_for_HBF}. In our
scheme, the entire vertical angle range, typically constrained within
a certain prior interval, is uniformly divided into $G$ sub-intervals.
Correspondingly, the $N_{y}$ columns in the UPA are divided into
$G$ groups, and each group is assigned a narrow beam pointing toward
a specific vertical angle sub-interval, as illustrated in Fig. \ref{fig:group-wise}.
The set of narrow beams is designed based on the filter design theory.
As a result, the proposed beam design resolves vertical angle estimation
ambiguity by ensuring different responses from different vertical
angles. Moreover, the group-wise narrow beam is naturally aligned
with the overall vertical angle range to exploit the available array
gains.
\item \textbf{Optimization of antenna grouping pattern for horizontal angle
estimation:} As illustrated in Fig. \ref{fig:group-wise}, due to
the group-wise narrow beam design, a channel path with the vertical
angle lying in one sub-interval may only have strong received signals
at the antennas in the corresponding group. As such, the effective
number of horizontal antennas used to estimate the horizontal angle
for this channel path is reduced, leading to a potential performance
loss in horizontal angle estimation. To mitigate this problem, we
further optimize the antenna grouping pattern to enhance horizontal
angle estimation accuracy. We introduce an optimization strategy that
jointly considers the integrated side-lobe level (ISL) {[}\cite{ISL_ref}{]}
and the Statistical Resolution Limit (SRL) \cite{SRL_ref} as performance
metrics in the horizontal dimension, where the ISL reflects interference
suppression capability and the SRL reflects resolution capability.
To efficiently address this challenging problem, we perform the Estimation
of Distribution Algorithm (EDA) \cite{EDA_ref} in an offline manner.
Such an optimization design achieves a better balance between resolution
and interference suppression capabilities, consequently, the overall
horizontal angle estimation performance is improved.
\item \textbf{Low-complexity group-wise SC-VBI (GW-SC-VBI) algorithm:} Building
upon the proposed group-wise beam design, we develop a low-complexity
group-wise SC-VBI algorithm. Specially, the received signals are partitioned
into multiple groups according to the range of vertical sub-intervals.
For each group, the SC-VBI algorithm is performed based on the corresponding
received signals and the associated angle grids. In this way, angle
grids for each group are non-overlapping and adjusted independently.
In the final few iterations, a joint estimation is conducted to enhance
angle estimation performance, particularly for angles that lie between
the boundaries of two adjacent groups. This group-wise approach reduces
the signal processing dimensions and significantly lowers the computational
complexity, making our scheme more suitable for practical PC-HBF systems,
which are typically equipped thousands of antennas.
\end{itemize}
The remainder of this paper is organized as follows. In Section \ref{sec:System-Model},
we present the system model and signal model. In Section \ref{sec:beam_design_method},
we introduce the group-wise narrow beam design in the vertical dimension.
In Section \ref{sec:grouping_pattern_opt}, we detail the optimization
of antenna grouping pattern using EDA in the horizontal dimension.
In Section \ref{sec:grouping_pattern_opt}, we introduce the proposed
low-complexity GW-SC-VBI algorithm. Finally, simulation results are
provided in Section \ref{sec:simulations} and conclusions are drawn
in Section \ref{sec:Conclusion}.

\textit{Notation:} $\left(\cdot\right)^{-1}$, $\left(\cdot\right)^{T}$,
$\left(\cdot\right)^{H}$, and $\left\Vert \cdot\right\Vert $ are
used to represent the inverse, transpose, conjugate transpose, and
$\ell_{2}\textrm{-norm}$, respectively. $\mathfrak{Re}\left\{ \cdot\right\} $
denotes the real part of the complex argument. For a vector $\boldsymbol{x}\in\mathbb{C}^{N}$
and a given index set $\mathcal{S}\subseteq\left\{ 1,...,N\right\} $,
$\left|\mathcal{S}\right|$ denotes its cardinality, $\boldsymbol{x}_{\mathcal{S}}\in\mathbb{C}^{\left|\mathcal{S}\right|\times1}$
denotes the subvector consisting of the elements of $\boldsymbol{x}$
indexed by the set $\mathcal{S}$. $\text{diag}\left(\boldsymbol{x}\right)$
denotes a block diagonal matrix with $\boldsymbol{x}$ as the diagonal
elements. $\otimes$ and $\odot$ denote the Kronecker product and
Hadamard product, respectively. $\mathcal{CN}\left(\boldsymbol{x};\boldsymbol{\mu},\mathbf{\Sigma}\right)$
represents a complex Gaussian distribution with mean $\boldsymbol{\mu}$
and covariance matrix $\mathbf{\Sigma}$. $\Gamma\left(x;a,b\right)$
represents a Gamma distribution with shape parameter $a$ and rate
parameter $b$. Finally, $x=\Theta(a)$ for $a>0$ denotes that $\exists k_{1},k_{2}>0$,
such that $k_{2}\cdot a\leq x\leq k_{1}\cdot a$.

\section{Symtem and Signal Model\label{sec:System-Model}}

\subsection{System Model}

We consider a narrowband massive MIMO system with a PC-HBF structure.
During the channel estimation stage, each user transmits uplink pilot
signals, and the BS estimates the channel based on the received signals,
as shown in Fig. \ref{fig:BS_height}. The BS is equipped with $N_{y}$
horizontal antennas and $N_{z}$ vertical antennas, leading to a total
of $N_{r}=N_{y}N_{z}$ antennas. Each RF chain is connected to a vertical
subarray consisting of $M$ antennas, where $M$ is also referred
to as the compression ratio. Consequently, the number of RF chains
is given by $N_{RF}=\frac{N_{r}}{M}$, as illustrated in Fig. \ref{fig:HBF_structure}.

Without loss of generality, we focus on a single-user scenario where
the user is equipped with a single antenna. However, the proposed
beam design and channel estimation algorithm can be easily extended
to a multi-user scenario where each user is equipped with multiple
antennas, since the uplink pilot signals transmitted by different
users and antennas are orthogonal during the channel estimation stage.

\subsection{Signal Model}

The received signal $\boldsymbol{y}\in\mathbb{C}^{N_{RF}\times1}$
at the BS can be written as:
\begin{equation}
\boldsymbol{y}=\mathbf{F}_{a}\boldsymbol{h}+\boldsymbol{w},\label{eq:HBF_signal_model}
\end{equation}
where $\boldsymbol{h}\in\mathbb{C}^{N_{r}\times1}$ is the channel
vector of the user, $\mathbf{F}_{a}\in\mathbb{C}^{N_{RF}\times N_{r}}$
is the analog beam matrix, and $\boldsymbol{w}\sim\mathcal{CN}(0,\sigma^{2}\textbf{I})\in\mathbb{C}^{N_{RF}\times1}$
is the additive white Gaussian Noise (AWGN). Note that we ignore the
digital beam matrix in this paper, as it is a full-rank matrix that
can be incorporated into the observations and does not affect the
channel estimation phase. In the PC-HBF structure, the analog beam
matrix is composed of a series of low-cost phase shifters, and the
non-zero elements of \textbf{$\mathbf{F}_{a}$} have unit modulus
constraint. Moreover, each RF chain is only connected to a vertical
subarray consisting of $M$ antennas. As a result, $\mathbf{F}_{a}$
is a block diagonal matrix, where each block corresponds to a row
vector with $M$ elements. The overall analog beam matrix $\mathbf{F}_{a}$
can be represented as the block diagonal concatenation of multiple
compression vectors:
\begin{equation}
\mathbf{F}_{a}=\textrm{blkdiag}\left(\boldsymbol{v}_{a}^{1},\ldots,\boldsymbol{v}_{a}^{N_{RF}}\right).\label{eq:compress_vector}
\end{equation}
where compression vector $\boldsymbol{v}_{a}^{k}\in\mathbb{C}^{1\times M}$
is the row vector corresponding to the $k$-th RF chain.

The channel vector $\boldsymbol{h}$ can be represented as:
\begin{equation}
\ensuremath{\boldsymbol{h}=\sum_{k=1}^{K}\alpha_{k}\boldsymbol{a}_{R}\left(\theta_{k},\varphi_{k}\right),}\label{eq:h_model}
\end{equation}
where $K$ is the number of propagation paths, $\alpha_{k}$, $\theta_{k}$
and $\varphi_{k}$ are the equivalent complex gain, azimuth angle
of arrival (AoA), and elevation AoA of the $k$-th channel path, respectively,
and $\boldsymbol{a}_{R}\left(\theta_{k},\varphi_{k}\right)\in\mathbb{C}^{N_{r}\times1}$
is the array response vector of UPA lying in the $yz$-plane, which
can be represented as $\mathbf{a}_{R}\left(\theta,\phi\right)=\mathbf{a}_{y}\left(\textrm{sin}\left(\theta\right),\textrm{cos}\left(\varphi\right)\right)\otimes\mathbf{a}_{z}\left(\textrm{sin}\left(\varphi\right)\right)$.
Specially, with half-wavelength spacing, the $n_{y}$-th element in
the steering vector $\mathbf{a}_{y}\left(\textrm{sin}\left(\theta\right),\textrm{cos}\left(\varphi\right)\right)$
and the $n_{z}$-th element in the steering vector $\mathbf{a}_{z}\left(\textrm{sin}\left(\varphi\right)\right)$
can be respectively expressed as $\left[\mathbf{a}_{y}\left(\textrm{sin}\left(\theta\right),\textrm{cos}\left(\varphi\right)\right)\right]_{n_{y}}=e^{j\pi n_{y}\textrm{sin}\left(\theta\right)\textrm{cos}\left(\varphi\right)}$,
$\left[\mathbf{a}_{z}\left(\textrm{sin}\left(\varphi\right)\right)\right]_{n_{z}}=e^{j\pi n_{z}\textrm{sin}\left(\varphi\right)}$,
with $n_{y}=0,\ldots,N_{y}-1$ and $n_{z}=0,\ldots,N_{z}-1$.

\section{Group-wise Narrow Beam Design in the Vertical Dimension \label{sec:beam_design_method}}

In this section, we first analyze the limitations of traditional beam
design approaches from the perspective of the ambiguity function (AF)
and highlight the challenges, which motivate the proposed group-wise
narrow beam design. Subsequently, we introduce the special beam design
criteria based on bandpass filter banks, which can effectively suppress
interference from undesired angular components.

\subsection{Limitation Analysis of Traditional Beam Design}

One of the most common beam schemes is the random beam employed in
\cite{SC_VBI}. However, the random beam is not sufficiently optimized
and does not fully exploit the available array gain, which significantly
limits its performance. An intuitive approach to better exploit the
array gain is to design all compression vectors to be identical wide
beams, acting as a bandpass filter that covers the entire vertical
angle range. Nevertheless, directly adopting such a wide beam results
in extremely poor channel estimation performance due to vertical ambiguity,
as described below from the perspective of the AF.

The AF represents the difficulty of identifying a specific vertical
angle component based on a simple matched filtering approach, providing
a fundamental metric on the performance of angle estimation \cite{MIMO_radar_AF,bistatic_radar_AF}.
In the vertical dimension, for a given analog beam matrix $\mathbf{F}_{a}$
and horizontal angle $\theta_{0}$, the ambiguity function in the
vertical dimension with respect to angle $\varphi_{1}$ and $\varphi_{2}$
can be expressed as:
\begin{equation}
\begin{aligned}\chi^{\textrm{ver}}\left(\varphi_{1},\varphi_{2}\mid\theta_{0}\right)= & \left(\mathbf{F}_{a}\boldsymbol{a}_{R}\left(\theta_{0},\varphi_{1}\right)\right)^{H}\mathbf{F}_{a}\boldsymbol{a}_{R}\left(\theta_{0},\varphi_{2}\right)\\
= & \boldsymbol{a}_{R}\left(\theta_{0},\varphi_{1}\right)^{H}\left(\mathbf{F}_{a}^{H}\mathbf{F}_{a}\right)\boldsymbol{a}_{R}\left(\theta_{0},\varphi_{2}\right).
\end{aligned}
\label{eq:AF_in_vertical}
\end{equation}

\begin{figure}
\begin{centering}
\includegraphics[width=80mm]{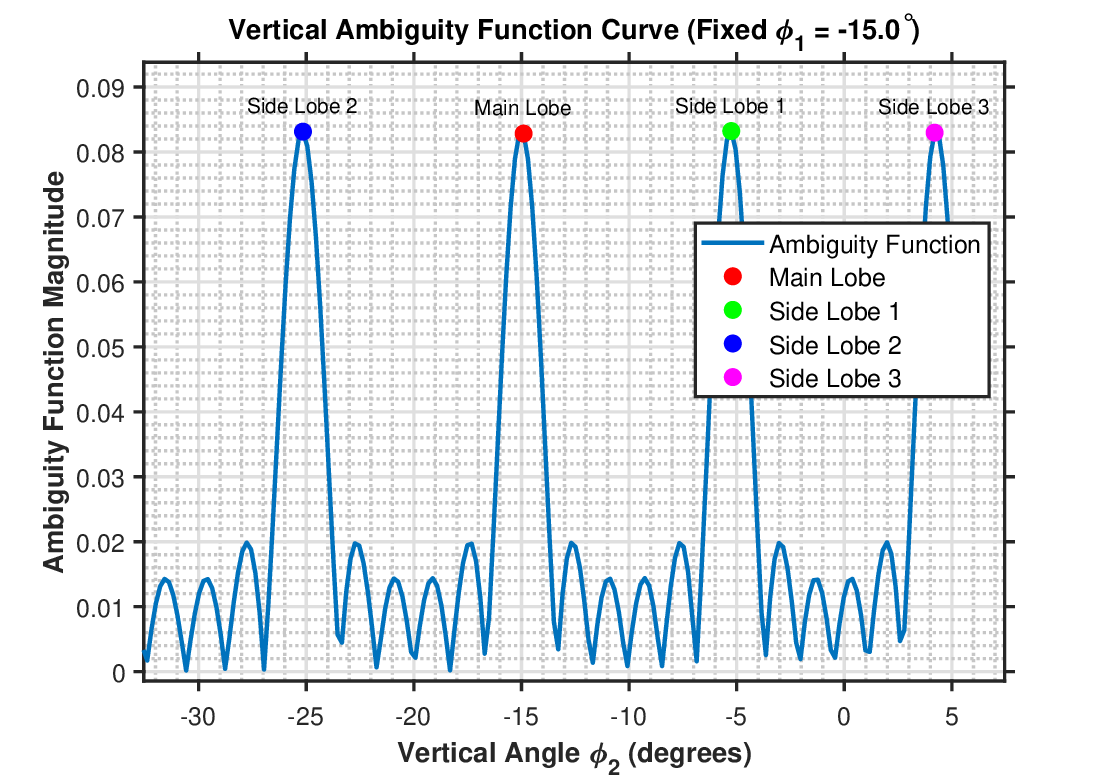}
\par\end{centering}
\caption{{\small\label{fig:AF_with_wide_beam}}Vertical AF when the wide beam
is adopted.}
\end{figure}

Note that unlike the traditional AF in \cite{MIMO_radar_AF,bistatic_radar_AF},
which is stationary and depends only on the difference between two
variables, the AF in \ref{eq:AF_in_vertical} depends on both $\varphi_{2}$
and $\varphi_{1}$, rather than solely on their difference $\varphi_{2}-\varphi_{1}$.
A desirable AF should satisfy the following condition: For any given
vertical angle $\varphi_{1}$, the AF with respect to the vertical
angle $\varphi_{2}$ should exhibit a single main-lobe at $\varphi_{2}=\varphi_{1}$
with the side-lobe portions occupying relatively less energy, which
ensures that any vertical angle does not cause significant interference
to other vertical angles, such that each vertical angle can be uniquely
identified. However, when the wide beam is adopted, the corresponding
AF, as shown in Fig. \ref{fig:AF_with_wide_beam}, exhibits multiple
side lobes with non-negligible energy compared to that of the main
lobe, resulting in significantly degraded vertical angle estimation
performance. This phenomenon is also called ambiguity in traditional
beamspace signal processing \cite{2005_avoid_ambiguity,2D_beamspace_MUSIC}.

In summary, the core challenge in vertical angle estimation is to
exploit array gain while ensuring that the designed beam is free from
ambiguity in the vertical dimension. To tackle this challenge, we
propose a group-wise narrow beam design using bandpass filter banks
in the next subsection.

\subsection{Group-wise Narrow Beam Design Using Bandpass Filter Banks}

Our primary goal is to design beams appropriately to suppress the
interference between the estimation of different vertical angles.
One potential approach is to directly optimize the analog beam matrix
$\mathbf{F}_{a}$ to minimize interference. However, due to the non-stationary
nature of the AF in the vertical dimension, this optimization problem
is very challenging.

The bandpass filter banks approach adopted in the CBS algorithm \cite{2020_tsp_CBS,2024_tsp_CBS_for_HBF}
offers a novel perspective for beam design. Specifically, by designing
multiple non-overlapping bandpass filter banks, the entire angle range
can be divided into smaller sub-intervals, with each sub-interval
processed in parallel across multiple timeslots. This effectively
attenuates signals from undesired sub-intervals and significantly
improves the vertical angle estimation accuracy. 

However, as mentioned earlier, each user is usually allocated a single
pilot symbol within one time slot, and channel estimation should be
completed within that single timeslot. This constraint makes it impractical
to design the beams directly based on the CBS approach in \cite{2020_tsp_CBS,2024_tsp_CBS_for_HBF}.
Fortunately, for vertical angle estimation, the $N_{y}$ columns of
the uniform planar array can be treated as independent observations.
This enables the total horizontal antennas to be naturally divided
into multiple groups, and the total vertical angle range is divided
into multiple sub-intervals correspondingly. 

In this way, each group is assigned a narrow beam pointing toward
a specific vertical angle sub-interval to generate observations for
recovering the vertical angles in this sub-interval. As a result,
all vertical angles in the vertical angle range can be recovered without
ambiguity, since under the proposed group-wise narrow beam, different
vertical angles will surely produce different response after analog
compression. Furthermore, this method allows the union of the analog
beams in all groups to naturally align with the vertical angle range
without additional signal processing, ensuring accurate vertical angle
estimation without loss of array gain.

We now present the proposed beam design scheme in detail. In a UPA
with $N_{y}$ columns and each column has $T=\frac{N_{z}}{M}$ RF
chains, the overall analog beam matrix $\mathbf{F}_{a}$ can be represented
as the block diagonal matrix: 
\begin{equation}
\mathbf{F}_{a}=\textrm{blkdiag}\left(\widehat{\mathbf{F}}_{a}^{1},\ldots,\widehat{\mathbf{F}}_{a}^{N_{y}}\right),
\end{equation}
where $\widehat{\mathbf{F}}_{a}^{i}\in\mathbb{C}^{T\times N_{z}}$
represents the analog beam sub-matrix corresponding to the $i$-th
column of the UPA.

We assume that the $N_{y}$ columns are divided into $G$ groups,
and the entire vertical angle range is partitioned into $G$ uniformly
spaced sub-intervals accordingly. It is worth noting that the uniformity
is achieved in the $\textrm{sin}\left(\varphi\right)$ rather than
directly with $\varphi$, consistent with the steering vector\textquoteright s
form. 

To represent antenna grouping pattern, we introduce binary grouping
vectors $\boldsymbol{s}_{g}\in\left\{ 0,1\right\} ^{N_{y}\times1}$
for $g=1,\ldots,G$, the nonzero indices in $\boldsymbol{s}_{g}$
indicate the column indices belonging to the $g$-th group. And there
is a constraint $\sum_{g=1}^{G}\mathbf{s}_{g}\leq\boldsymbol{1}$,
which ensures that all columns of the UPA array are allocated to different
groups without overlapping. For antennas in $g$-th group, a narrow
beam pointing to the sub-interval $\Phi_{g}$ is adopted and denoted
by $\mathbf{W}_{g}\in\mathbb{C}^{T\times N_{z}}$. We define $\textrm{supp}\left(\cdotp\right)$
as the set of indices corresponding to the nonzero entries in the
vector, such that for any $i\in\textrm{supp}\left(\boldsymbol{s}_{g}\right)$,
we have $\widehat{\mathbf{F}}_{a}^{i}=\mathbf{W}_{g}$, which acts
a narrow bandpass filter covering the sub-interval $\Phi_{g}$, such
that overall analog beam matrix can be written as:
\begin{equation}
\mathbf{F}_{a}=\sum_{g=1}^{G}\mathbf{W}_{g}\otimes\textrm{diag}\left(\boldsymbol{s}_{g}\right)\in\mathbb{C}^{N_{RF}\times N_{r}}.\label{eq:GW_narrow_beam}
\end{equation}

Since the horizontal antenna grouping pattern has little effect on
the performance of vertical angle estimation, we fix $\boldsymbol{s}_{g}$
for $g=1,\ldots,G$ in this section and focus the design on $\mathbf{W}_{g}$,
$g=1,\ldots,G$.

In the PC-HBF structure, the matrix $\mathbf{W}_{g}$ is constrained
to have constant-modulus entries and only a subset of its elements
are nonzero. In fact, $\mathbf{W}_{g}$ is formed by the block diagonal
concatenation of $T$ compression vectors, which can be expressed
as:
\begin{equation}
\mathbf{W}_{g}=\textrm{blkdiag}\left(\boldsymbol{v}_{g}^{\left(1\right)},\ldots,\boldsymbol{v}_{g}^{\left(T\right)}\right),
\end{equation}
where each $\boldsymbol{v}_{g}^{\left(t\right)}\in\mathbb{C}^{1\times M}$
is a constant-modulus vector. By leveraging the filter design theory,
each compression vector is identically designed to act as a narrow
bandpass filter covering the sub-interval $\Phi_{g}$, such that the
entire matrix $\mathbf{W}_{g}$ naturally functions as a narrow bandpass
filter for $\Phi_{g}$ and can be further simplified as:
\begin{equation}
\mathbf{W}_{g}=\textrm{blkdiag}\left(\mathbf{a}_{z}\left(\omega_{g}\right),\ldots,\mathbf{a}_{z}\left(\omega_{g}\right)\right),
\end{equation}
where $\mathbf{a}_{z}\left(\omega_{g}\right)\in\mathbb{C}^{1\times M}$
satisfies the constant-modulus constraint and represents the designed
narrow beam pointing to the center angle $\omega_{g}$ of the sub-interval
$\Phi_{g}$.

It is noteworthy that the CBS algorithm proposed in \cite{2020_tsp_CBS,2024_tsp_CBS_for_HBF}
cannot be directly applied to this scenario because deploying CBS
within the HBF structure necessitates a fully connected structure,
whereas the current framework employs a partially connected structure.
Moreover, our proposed beam design scheme offers additional advantages
over the CBS approach. The CBS scheme combines filters with a decimator,
with the number of required RF chains determined by both the filter
length and the decimation ratio, thereby imposing limitations on the
practical flexibility of the CBS algorithm. In contrast, our design
is compatible with any partially connected structure, offering greater
deployment freedom in real-world systems.

\section{Antenna Grouping Pattern Optimization in the Horizontal Dimension\label{sec:grouping_pattern_opt}}

As explained in the introduction, the above group-wise narrow beam
design reduces the effective number of horizontal antennas used for
estimating the horizontal angle of a specific channel path, which
may lead to performance degradation in horizontal angle estimation.
To mitigate this issue, we propose an optimization strategy for the
antenna grouping pattern in the horizontal dimension to offer a balanced
trade-off between resolution and interference suppression capability,
ultimately improving the overall performance in horizontal angle estimation.

\subsection{Performance Metrics}

To optimize the antenna grouping pattern, we consider two key performance
metrics: Integrated Side-Lobe Level (ISL) and Statistical Resolution
Limit (SRL).

\subsubsection{Horizontal ISL}

We first define the AF in the horizontal dimension. Similar to the
vertical case, for a given analog beam matrix $\mathbf{F}_{a}$ and
vertical angle $\varphi_{0}$, the conventional AF with respect to
horizontal angles $\theta_{1}$ and $\theta_{2}$ is given by:
\begin{equation}
\begin{aligned}\chi\left(\theta_{1},\theta_{2}\mid\varphi_{0}\right)= & \left(\mathbf{F}_{a}\boldsymbol{a}_{R}\left(\theta_{1},\varphi_{0}\right)\right)^{H}\mathbf{F}_{a}\boldsymbol{a}_{R}\left(\theta_{2},\varphi_{0}\right)\\
= & \boldsymbol{a}_{R}\left(\theta_{1},\varphi_{0}\right)^{H}\left(\mathbf{F}_{a}^{H}\mathbf{F}_{a}\right)\boldsymbol{a}_{R}\left(\theta_{2},\varphi_{2}\right).
\end{aligned}
\end{equation}

In our group-wise approach, the analog beam matrix for the $g$-th
group can be represented as $\mathbf{F}_{g}=\mathbf{W}_{g}\otimes\textrm{diag}\left(\boldsymbol{s}_{g}\right)$.
As a result, the horizontal AF for the $g$-th group is defined as:
\begin{equation}
\begin{aligned}\chi_{g}\left(\theta_{1},\theta_{2}\mid\varphi_{g}\right)= & \left(\mathbf{F}_{g}\boldsymbol{a}_{R}\left(\theta_{1},\varphi_{g}\right)\right)^{H}\mathbf{F}_{g}\boldsymbol{a}_{R}\left(\theta_{2},\varphi_{g}\right)\\
= & \boldsymbol{a}_{R}\left(\theta_{1},\varphi_{g}\right)^{H}\left(\mathbf{\mathbf{F}}_{g}^{H}\mathbf{F}_{g}\right)\boldsymbol{a}_{R}\left(\theta_{2},\varphi_{g}\right)\\
= & \boldsymbol{a}_{R}\left(\theta_{1},\varphi_{g}\right)^{H}\left[\left(\mathbf{W}_{g}^{H}\mathbf{W}_{g}\right)\otimes\textrm{diag}\left(\boldsymbol{s}_{g}\right)\right]\boldsymbol{a}_{R}\left(\theta_{2},\varphi_{g}\right)\\
= & \boldsymbol{s}_{g}^{T}\mathbf{a}_{y}\left(\Delta,\varphi_{g}\right)M,
\end{aligned}
\label{eq:horizontal_AF}
\end{equation}
where $\boldsymbol{s}_{g}$ represents the binary grouping vector
for the $g$-th group, $\varphi_{g}$ is the central angle of the
sub-interval $\Phi_{g}$, $\Delta=\textrm{sin}\left(\theta_{2}\right)-\textrm{sin}\left(\theta_{1}\right)$,
and $M=\mathbf{a}_{z}\left(\varphi_{g}\right)^{H}\mathbf{W}_{g}^{H}\mathbf{W}_{g}\mathbf{a}_{z}\left(\varphi_{g}\right)$
is a constant independent of $\Delta$. This formulation evaluates
the AF on a per-group basis rather than the overall AF, thereby reflecting
the individual interference suppression performance of each group. 

Note that the AF in the horizontal dimension is stationary, i.e, it
depends solely on $\Delta$ rather than on the absolute angles, since
no compression is performed in the horizontal dimension. The stationary
behavior in the horizontal dimension facilitates more straightforward
analysis and design.

For convenience, we omit $\varphi_{g}$ and define $\chi_{g}\left(\Delta\right)\triangleq\chi_{g}\left(\theta_{1},\theta_{2}\mid\varphi_{g}\right)$,
and the horizontal ISL for the $g$-th group is defined as:
\begin{equation}
\textrm{ISL}_{g}\triangleq\frac{\int_{\mathcal{R}_{s}}\left|\chi_{g}\left(\Delta\right)\right|^{2}d\Delta}{\left|\mathcal{R}_{s}\right|\left|\chi_{g}\left(0\right)\right|^{2}},\label{eq:ISL_definition}
\end{equation}
where $\mathcal{R}_{s}$ is a symmetric interval representing the
sidelobe region of the horizontal ISL. Substituting (\ref{eq:horizontal_AF})
into (\ref{eq:ISL_definition}), the horizontal ISL can be further
expressed as:
\begin{equation}
\begin{aligned}\textrm{ISL}_{g}= & \frac{\boldsymbol{s}_{g}^{T}\mathbf{V}_{g}\boldsymbol{s}_{g}}{\left|\mathcal{R}_{s}\right|\left(\mathbf{s}_{g}^{T}\boldsymbol{1}\right)^{2}}\end{aligned}
,\label{eq:ISL_simplified}
\end{equation}
where $\boldsymbol{1}$ represents the vector composed of $1$, and
$\mathbf{V}_{g}=\int_{\mathcal{R}_{s}}\mathbf{a}_{y}\left(\Delta,\varphi_{g}\right)\mathbf{a}_{y}\left(\Delta,\varphi_{g}\right)^{H}d\Delta$
is a symmetric Toeplitz matrix, which can be fully determined by its
first column $\boldsymbol{f}_{g}\left(n\right)$:
\begin{equation}
\boldsymbol{f}_{g}\left(n\right)=\begin{cases}
\frac{2\left(\textrm{sin}\left(n\pi\textrm{cos}\left(\varphi_{g}\right)b\right)-\textrm{sin}\left(n\pi\textrm{cos}\left(\varphi_{g}\right)a\right)\right)}{n\pi\textrm{cos}\left(\varphi_{g}\right)} & n=1,\ldots,N_{y}-1,\\
2b-2a & n=0.
\end{cases}
\end{equation}

Note that the sidelobe of one angle will cause interference to the
estimation of other angles. Since the horizontal ISL\textcolor{blue}{{}
}in (\ref{eq:ISL_definition})\textcolor{black}{{} represents the relative
amount of energy in side-lobe portions with respect to the main-lobe
portions}, the horizontal ISL evaluates the interference suppression
capability under a certain antenna grouping pattern. Generally, the
uniform grouping pattern achieves a very small ISL value while the
ISL of the random grouping pattern is relatively large.

\subsubsection{Horizontal SRL}

The group-wise narrow beam design may also reduces the effective antenna
aperture in the horizontal if the antennas in a group concentrate
on a small range such as that in the uniform grouping pattern. Consequently,
both side-lobe interference and resolution capability need to be considered
when evaluating horizontal beam design. To evaluate resolution capability,
we further introduce horizontal angle Statistical Resolution Limit
(SRL), which reflects the system\textquoteright s ability to resolve
closely spaced horizontal angles.

The horizontal angle SRL is defined as \cite{SRL_ref}:
\begin{eqnarray}
 & \mathrm{SRL} & \triangleq\Delta\nonumber \\
\text{\textrm{s.t.}} & \Delta & =\sqrt{\textrm{CRB}_{\Delta}},\label{eq:SRL_definition}
\end{eqnarray}
where $\textrm{CRB}_{\Delta}$ denotes the Cram\'er-Rao bound (CRB)
for the horizontal angle separation $\Delta=\textrm{sin}\left(\theta_{2}\right)-\textrm{sin}\left(\theta_{1}\right)$,
and serves as a lower bound on the mean-squared error (MSE) achieved
by any unbiased estimator \cite{estimation_theory_book}. When the
standard deviation of the estimated horizontal angle separation equals
the true separation, resolution limit is achieved. Therefore, the
SRL establishes the minimum horizontal angle separation that can be
reliably resolved, providing a fundamental limit on the resolution
capability of the antenna grouping pattern, i.e., a smaller SRL indicates
enhanced resolution performance.

Now we analyze the SRL for the $g$-th group. The CRB for the horizontal
angle separation within the $g$-th group, denoted as $\textrm{CRB}_{\Delta_{g}}$,
is given by:
\begin{eqnarray}
\textrm{CRB}_{\Delta_{g}} & = & \frac{\partial\Delta_{g}}{\partial\boldsymbol{\mu}_{g}}^{T}\mathbf{J}_{\boldsymbol{\mu}_{g}}^{-1}\frac{\partial\Delta_{g}}{\partial\boldsymbol{\mu}_{g}},\nonumber \\
 & = & \mathbf{J}_{\boldsymbol{\mu}_{g}}^{-1}(1,1)+\mathbf{J}_{\boldsymbol{\mu}_{g}}^{-1}(2,2)-\mathbf{J}_{\boldsymbol{\mu}_{g}}^{-1}(1,2)\nonumber \\
 &  & -\mathbf{J}_{\boldsymbol{\mu}_{g}}^{-1}(2,1),
\end{eqnarray}
where $\Delta_{g}=\textrm{sin}\left(\theta_{2,g}\right)-\textrm{sin}\left(\theta_{1,g}\right)$,
$\mathbf{J}_{\boldsymbol{\mu}_{g}}\triangleq\mathbb{E}_{\boldsymbol{y}}\left[-\frac{\partial^{2}\ln\mathit{f}(\boldsymbol{y}|\boldsymbol{\mu}_{g})}{\partial\boldsymbol{\mu}_{g}\partial\boldsymbol{\mu}_{g}^{T}}\right]$
is the Fisher information matrix (FIM) associated with the unknown
parameter vector $\boldsymbol{\mu}_{g}$, $f(\boldsymbol{y}|\boldsymbol{\mu}_{g})$
is the likelihood function of the random vector $\boldsymbol{y}$
conditioned on $\boldsymbol{\mu}_{g}$, and $\boldsymbol{\mu}_{g}\triangleq[\boldsymbol{\beta}_{g}^{T},\boldsymbol{\alpha}_{g}^{T}]$
can be represented as:
\begin{equation}
\begin{aligned} & \boldsymbol{\beta}_{g}=\left[\textrm{sin}\left(\theta_{1,g}\right),\textrm{sin}\left(\theta_{2,g}\right)\right]^{T},\\
 & \boldsymbol{\alpha}_{g}=\left[\alpha_{1,g}^{R},\alpha_{2,g}^{R},\alpha_{1,g}^{I},\alpha_{2,g}^{I}\right]^{T},
\end{aligned}
\end{equation}
where $\alpha_{1,g}^{R}$ and $\alpha_{1,g}^{I}$ denote the real
and imaginary parts of channel gain $\alpha{}_{1,g}$, respectively,
and the same applies to $\alpha_{2,g}^{R}$ and $\alpha_{2,g}^{I}$.
Detailed derivations of $\mathbf{J}_{\boldsymbol{\mu}_{g}}$ can be
found in Appendix \ref{subsec:AppexA}.

It is important to note that the random grouping pattern generally
achieves a smaller SRL than the uniform grouping pattern, owing to
its larger effective aperture, which enhances resolution. In contrast,
the uniform grouping pattern generally results in a larger SRL due
to its limited effective aperture, thereby restricting resolution
capability.

\subsection{Problem Formulation \label{subsec:Problem-Formulation}}

The above observations highlight an inherent trade-off: while a random
pattern improves resolution at the expense of increased interference,
a uniform pattern reduces interference at the cost of sacrificing
resolution. This trade-off motivates the need for an optimized antenna
grouping pattern that can strike a balance between minimizing ISL
and achieving a low SRL, thereby enhancing the overall horizontal
angle estimation performance.

Accordingly, our objective is to minimize the maximum ISL among all
groups while ensuring that each group maintains a sufficiently small
SRL as:
\begin{eqnarray}
\mathcal{P}: & \underset{\mathcal{X}}{\min} & \underset{g}{\max}\left\{ \textrm{ISL}_{g}\right\} \nonumber \\
 & \text{s.t. } & \ensuremath{\textrm{SRL}_{g}\leq\rho_{g},\forall g},\label{eq:SRL_max}\\
 &  & \sum_{g=1}^{G}\mathbf{s}_{g}\leq\boldsymbol{1},\label{eq:non_onverlap_constraint}\\
 &  & \mathbf{s}_{g}(n)\in\left\{ 0,1\right\} ,\forall n,
\end{eqnarray}
where $\mathcal{X}\triangleq\left\{ \mathbf{s}_{1},...,\mathbf{s}_{G}\right\} $
denotes the set of optimization variables, $\mathbf{s}_{g}(n)$ represents
the $n$-th element of $\mathbf{s}_{g}$, $\textrm{SRL}_{g}$ denotes
the SRL of the $g$-th group, and $\rho_{g}$ denotes the upper bound
imposed on $\textrm{SRL}_{g}$. Constraint (\ref{eq:SRL_max}) guarantees
sufficient horizontal angle resolution. Constraint (\ref{eq:non_onverlap_constraint})
ensures that all columns of the UPA array are allocated to different
groups without overlapping. Consequently, the optimized grouping pattern
is expected to reach an enhanced horizontal angle estimation performance.

It is important to note that $\text{\ensuremath{\textrm{SRL}_{g}}}$
is influenced by the noise variance and path gains of the system during
the FIM calculation. To facilitate offline computation of each group\textquoteright s
SRL, we assume fixed values for noise variance $\sigma_{g}^{2}$ ,
vertical angle $\varphi_{g}$, and the path gains $\alpha_{1,g},\alpha_{2,g}$
for $g=1,\ldots,G$, so that SRL of all groups can be calculated in
an offline manner. 

In the simulations, we set $\alpha_{1,g}=\alpha_{2,g}=1$, $\sigma_{g}=0.18$,
and $\varphi_{g}$ is set as the central angle of the sub-interval
$\Phi_{g}$, $\forall g$. Note that the setting of these parameters
have relatively little effect on the final performance, since we can
adjust the corresponding $\rho_{g}$ to ensure that the SRL constraint
(\ref{eq:SRL_max}) is always effective. For given $\sigma_{g}^{2}$,
$\varphi_{g}$ and $\alpha_{1,g},\alpha_{2,g}$, we calculate the
horizontal SRL value under a random grouping pattern as a reference
for setting $\rho_{g}$. This is motivated by the fact that the random
grouping pattern typically achieves a smaller SRL than other grouping
patterns, as it tends to span a wider frequency range with high probability.
Moreover, the specific choice of these parameters has relatively little
effect on the final horizontal angle estimation performance, as validated
in section \ref{sec:simulations}.

The optimization of $\mathcal{P}$ is an integer nonlinear programming
(INLP) problem, which is non-deterministic polynomial-time hard (NP-hard).
In next subsection, we will employ the EDA method to obtain an efficient
sub-optimal solution for this problem.

\subsection{EDA Algorithm}

EDA is an evolutionary algorithm that operates by learning and sampling
from the probability distribution of the best individuals in each
generation. This approach allows EDA to exhibit superior global search
capabilities and robustness compared to traditional evolutionary algorithms
when addressing INLP problems \cite{EDA_ref}. The EDA process can
be broken down into several key steps:
\begin{itemize}
\item \textbf{Random Initialization}
\end{itemize}
EDA begins with the random initialization of a population $\mathcal{S}^{\left(0\right)}$
consisting of $Q$ individuals, i.e., $\mathcal{S}^{\left(0\right)}=\{\mathbf{S}_{1}^{\left(0\right)},...,\mathbf{S}_{Q}^{\left(0\right)}\}$.
Each individual $\mathbf{S}_{q}^{\left(0\right)}\in\left\{ 0,1\right\} ^{N_{y}\times G}$
is a binary matrix representing the antenna grouping pattern matrix,
and each row of $\mathbf{S}_{q}^{\left(0\right)}$ has at most one
non-zero element in order to satisfy the non-overlapping constraint
(\ref{eq:non_onverlap_constraint}). Additionally, only those individuals
that meet the SRL constraint (\ref{eq:SRL_max}) are selected as candidates
for further processing.
\begin{itemize}
\item \textbf{Selection of Best Individuals}
\end{itemize}
In the $i$-th iteration, the algorithm identifies the best $T$ $(T<Q)$
individuals from the previous population $\mathcal{S}^{\left(i-1\right)}$.
These selected individuals are denoted as $\mathcal{\widetilde{S}}^{\left(i\right)}=\{\widetilde{\mathbf{S}}_{1}^{\left(i\right)},...,\widetilde{\mathbf{S}}_{T}^{\left(i\right)}\}$.
The goodness of the individuals is measured by the objective function
value from problem $\mathcal{P}$, with a lower ISL value indicating
a more favorable individual.
\begin{itemize}
\item \textbf{Probability Calculation}
\end{itemize}
The EDA algorithm computes the probability of assigning the $n$-th
column of the UPA array to the $g$-th group by averaging among the
individuals in $\mathcal{\widetilde{S}}^{\left(i\right)}$, i.e.,
\begin{equation}
p^{\left(i\right)}(n,g)=\frac{1}{T}\sum_{t=1}^{T}\widetilde{\mathbf{S}}_{t}^{\left(i\right)}(n,g),\label{eq:Pro_in_EDA}
\end{equation}
and the total occurrence probability matrix is defined as $\mathbf{P}^{\left(i\right)}\in\mathbb{R}^{N_{y}\times G}$.
\begin{itemize}
\item \textbf{New Population Generation}
\end{itemize}
Using the calculated probabilities, a new population $\mathcal{S}^{\left(i\right)}=\{\mathbf{S}_{1}^{\left(i\right)},...,\mathbf{S}_{Q}^{\left(i\right)}\}$
is generated. For each individual $\mathbf{S}_{q}^{\left(i\right)}$,
the element $\mathbf{S}_{q}^{\left(i\right)}\left(n,g\right)$ is
determined according to the Bernoulli distribution:

\begin{equation}
\mathbf{S}_{q}^{\left(i\right)}\left(n,g\right)\sim\textrm{Bern}\left(p^{\left(i\right)}\left(n,g\right)\right),
\end{equation}
where $\textrm{Bern}\left(p^{\left(i\right)}\left(n,g\right)\right)$
represents a Bernoulli distribution with the probability $P\left(\mathbf{S}_{q}^{\left(i\right)}\left(n,g\right)=1\right)=p^{\left(i\right)}\left(n,g\right)$.
Moreover, this generation process is conducted under the condition
that all individuals satisfy constraint (\ref{eq:SRL_max})-(\ref{eq:non_onverlap_constraint}).
Finally, the best individual from the previous population $\mathcal{S}^{\left(i-1\right)}$
is retained in the new population $\mathcal{S}^{\left(i\right)}$.
The algorithm continues to iterate until it reaches the maximum number
of iterations $I_{max}$. 

The detailed convergence analysis is presented in \cite{EDA_convergence},
demonstrating that the EDA algorithm converges to the global optimum
as the number of individuals $Q$ approaches infinity. Note that we
adopt offline rather than online optimization, the optimization of
antenna grouping pattern based on the EDA algorithm will not introduce
any additional complexity for channel estimation. Finally, the EDA
algorithm for antenna grouping pattern optimization is summarized
in Algorithm \ref{alg:EDA_algorithm}.

\begin{algorithm}[t]
{\small\caption{\label{alg:EDA_algorithm}The EDA algorithm for antenna grouping pattern
optimization.}
}{\small\par}

\textbf{Input:} individual number parameters $Q,T$\textcolor{black}{,
maximum iteration number }$I_{max}$\textcolor{black}{, }ISL relevant
parameters $\mathbf{V}_{g},\mathcal{R}_{s}$, and SRL relevant parameters
$\beta_{g},\sigma_{g},\rho_{g,1},\rho_{g,2},\forall g$.

\textbf{Output:} The sub-optimal antenna grouping pattern matrix $\mathbf{S}_{\textrm{best}}^{\left(I_{max}\right)}$.

\begin{algorithmic}[1]

\STATE Randomly generate $Q$ individuals of the antenna grouping
pattern matrix $\mathbf{S}_{q}^{\left(0\right)}$ to initialize the
population $\mathcal{S}^{\left(0\right)}$.

\FOR{ $i=1,\cdots,I_{max}$}

\STATE Evaluate the fitness of individuals $\mathbf{S}_{q}^{\left(i-1\right)}$
in $\mathcal{S}^{\left(i-1\right)}$ by calculating the objective
function ISL.

\STATE Select the best $T$ individuals from $\mathcal{S}^{\left(i-1\right)}$
to construct $\mathcal{\widetilde{S}}^{(i)}$.

\STATE Calculate occurrence probability matrix $\mathbf{P}^{\left(i\right)}\in\mathbb{R}^{N\times G}$,
with each element $p^{\left(i\right)}\left(n,g\right)$ determined
by (\ref{eq:Pro_in_EDA}).

\STATE Construct the proper conditional Bernoulli probability distribution
$\textrm{Bern}\left(\mathbf{P}^{\left(i\right)}\right)$ with probability
$P\left(\mathbf{S}_{q}^{\left(i\right)}\left(n,g\right)=1\right)=p^{\left(i\right)}\left(n,g\right)$
subject to the constraints (\ref{eq:SRL_max})-(\ref{eq:non_onverlap_constraint}).

\STATE Sample new individuals $\mathbf{S}_{q}^{\left(i\right)}$
from probability distribution $\textrm{Bern}\left(\mathbf{P}^{\left(i\right)}\right)$,
i.e., $\mathbf{S}_{q}^{\left(i\right)}\left(n,g\right)\sim\textrm{Bern}\left(p^{\left(i\right)}\left(n,g\right)\right)$.

\STATE Generate a new population $\mathcal{S}^{\left(i\right)}$
with individuals $\mathbf{S}_{q}^{\left(i\right)},q=1,...,Q$.

\STATE Preserve the best individual $\mathbf{S}_{\textrm{best}}^{\left(i-1\right)}$
from $\mathcal{S}^{\left(i-1\right)}$ to $\mathcal{S}^{\left(i\right)}$,
i.e., $\mathbf{S}_{1}^{\left(i\right)}=\mathbf{S}_{\textrm{best}}^{\left(i-1\right)}$.

\ENDFOR

\end{algorithmic}
\end{algorithm}

\section{Low Complexity GW-SC-VBI Algorithm for Channel Estimation\label{sec:GW-SC-VBI}}

To further exploit the advantages of the proposed group-wise narrow
beam design, we introduce a low-complexity GW-SC-VBI algorithm for
channel estimation. We first establish a sparse channel representation
in the angle domain. Then we demonstrate how the group-wise narrow
beam decomposes the original large-scale CS problem into multiple
parallel smaller-scale sub-problems, enabling parallel processing
and significantly reducing computational complexity. Building upon
this decomposition, we propose the GW-SC-VBI algorithm, which applies
the SC-VBI method in \cite{SC_VBI} to each group for jointly angle
grids refinement and sparse signal estimation. Furthermore, to address
the limited out-of-band suppression of the band-pass filter and mitigate
interference from neighboring sub-intervals, a robust joint processing
design is incorporated into the group-wise estimation framework, thereby
further enhancing estimation performance.

\subsection{Angle Domain Sparse Representation of Channels}

Similar to the approach in \cite{wanyubo_extrapolation}, we introduce
a 2D dynamic angular-domain grid consisting of $L=L_{1}\times L_{2}$
angle points , where the $l\textrm{-th}$ grid point is denoted by
$\left(\theta_{l},\varphi_{l}\right)$. We define $\boldsymbol{\theta}\triangleq\left[\theta_{1},\ldots,\theta_{L}\right]^{T}$
and $\boldsymbol{\varphi}\triangleq\left[\varphi_{1},\ldots,\varphi_{L}\right]^{T}$
as the dynamic grid vectors, with the uniform angle grid serving as
the initial value for the dynamic grid.

With the definition of the dynamic angular-domain grid, we can reformulate
the received signal model in (\ref{eq:h_model}) as a CS model with
dynamic angle grids:
\begin{equation}
\boldsymbol{y}=\mathbf{F}_{a}\mathbf{A}\left(\boldsymbol{\varOmega}\right)\boldsymbol{x}+\boldsymbol{w},\label{eq:CS_model}
\end{equation}
where $\boldsymbol{\varOmega}\triangleq\left\{ \boldsymbol{\theta},\boldsymbol{\varphi}\right\} $
is the collection of grid parameters, $\mathbf{A}\left(\boldsymbol{\varOmega}\right)\triangleq\left[\boldsymbol{a}\left(\theta_{1},\varphi_{1}\right),\ldots,\boldsymbol{a}\left(\theta_{L},\varphi_{L}\right)\right]\in\mathbb{C}^{N_{r}\times L}$
serves as the basis matrix, $\mathbf{F}_{a}\mathbf{A}\left(\boldsymbol{\varOmega}\right)\in\mathbb{C}^{N_{RF}\times L}$
forms the sensing matrix of the CS problem, and $\boldsymbol{x}\in\mathbb{C}^{L\times1}$
represents the angular-domain sparse channel vector, which contains
only $K\ll L$ non-zero elements corresponding to $K$ paths. Specifically,
the $l$-th element of $\boldsymbol{x}$, denoted by $x_{l}$, represents
the complex gain of the channel path lying around the $l$-th angle
grid point.

\subsection{Parallel Processing with Group-wise Narrow Beam}

To fully take advantage of the proposed group-wise narrow beam design,
the basis matrix $\mathbf{A}\left(\boldsymbol{\varOmega}\right)$
are partitioned column-wise into $G$ blocks:
\begin{equation}
\mathbf{A}\left(\boldsymbol{\varOmega}\right)=\left[\mathbf{A}_{1}\left(\boldsymbol{\varOmega}_{1}\right),\ldots,\mathbf{A}_{G}\left(\boldsymbol{\varOmega}_{G}\right)\right],\label{eq:many_part_A}
\end{equation}
where the $g$-th block $\mathbf{A}_{g}\left(\boldsymbol{\varOmega}_{g}\right)\in\mathbb{C}^{N_{r}\times L_{g}}$
is the basis matrix constructed from the angle grid falling inside
the sub-interval $\Phi_{g}$, and $\sum_{g=1}^{G}L_{g}=L$. Accordingly,
the sparse angle-domain vector $\boldsymbol{x}$ can be expressed
as a concatenation of $G$ vectors:
\begin{equation}
\boldsymbol{x}=\left(\boldsymbol{x}_{1}^{T},\ldots,\boldsymbol{x}_{G}^{T}\right)^{T},\label{eq:many_part_x}
\end{equation}
where $\boldsymbol{x}_{g}\in\mathbb{C}^{L_{g}\times1}$ represents
the sparse vector corresponding to the $g$-th group.

Recall that the analog beam matrix for the $g$-th group is defined
as $\mathbf{F}_{g}=\mathbf{W}_{g}\otimes\textrm{diag}\left(\boldsymbol{s}_{g}\right)$
in (\ref{eq:horizontal_AF}). For convenience, we define the indices
corresponding to the nonzero rows of $\mathbf{F}_{g}$ as $\mathcal{R}_{g}$
and $R_{g}=\left|\mathcal{R}_{g}\right|$. Then according to (\ref{eq:many_part_A})
and (\ref{eq:many_part_x}), the equivalent observation model for
the $g$-th group can be written as:
\begin{equation}
\begin{aligned}\boldsymbol{y}_{g}= & \overline{\mathbf{F}}_{g}\mathbf{A}\left(\boldsymbol{\varOmega}\right)\boldsymbol{x}+\boldsymbol{w}_{g}\\
= & \sum_{g=1}^{G}\overline{\mathbf{F}}_{g}\mathbf{A}_{g}\left(\boldsymbol{\varOmega}_{g}\right)\boldsymbol{x}_{g}+\boldsymbol{w}_{g},
\end{aligned}
\label{eq:sum_G_model}
\end{equation}
where $\overline{\mathbf{F}}_{g}=\mathbf{F}_{g}\left(\mathcal{R}_{g},:\right)\in\mathbb{C}^{R_{g}\times N_{r}}$,
$\boldsymbol{y}_{g}=\boldsymbol{y}\left(\mathcal{R}_{g},:\right)$
is the corresponding received signal, and $\boldsymbol{w}_{g}=\boldsymbol{w}\left(\mathcal{R}_{g},:\right)$
represents the corresponding noise. 

Since $\overline{\mathbf{F}}_{g}$ can be viewed as the concatenation
of several narrow beams $\mathbf{W}_{g}$, which exhibit strong attenuation
for out-of-band angle components, the cross-term $\overline{\mathbf{F}}_{g}\mathbf{A}_{i}\left(\boldsymbol{\varOmega}_{i}\right),i\neq g$
can be approximated as zero, such that the model in (\ref{eq:sum_G_model})
simplifies to a low-dimensional CS problem:
\begin{equation}
\begin{aligned}\boldsymbol{y}_{g}\approx & \overline{\mathbf{F}}_{g}\mathbf{A}_{g}\left(\boldsymbol{\varOmega}_{g}\right)\boldsymbol{x}_{g}+\boldsymbol{w}_{g}\\
= & \mathbf{\Xi}_{g}\left(\boldsymbol{\varOmega}_{g}\right)\boldsymbol{x}_{g}+\boldsymbol{w}_{g}
\end{aligned}
,\label{eq:low_dim_CS_model}
\end{equation}
where $\mathbf{\Xi}_{g}\left(\boldsymbol{\varOmega}_{g}\right)=\overline{\mathbf{F}}_{g}\mathbf{A}_{g}\left(\boldsymbol{\varOmega}_{g}\right)$
is the sensing matrix. Note that for the $g$-th group, both the sensing
matrix $\mathbf{\Xi}_{g}\left(\boldsymbol{\varOmega}_{g}\right)\in\mathbb{C}^{R_{g}\times L_{g}}$
and the sparse signal $\boldsymbol{x}_{g}\in\mathbb{C}^{L_{g}\times1}$
have significantly smaller dimension compared to the original problem
in (\ref{eq:CS_model}). In summary, this group-wise decomposition
reduces the dimension of the original CS problem and enables parallel
processing for each group, as will be further detailed in the next
subsection.

\subsection{GW-SC-VBI algorithm}

\subsubsection{Outline of GW-SC-VBI Algorithm}

The overall framework of the GW-SC-VBI algorithm is divided into two
parts:
\begin{itemize}
\item \textbf{SC-VBI algorithm for each group: }The SC-VBI algorithm is
independently applied to each group and can be efficiently implemented
in parallel for all groups.
\item \textbf{Robust design based on joint processing:} After the SC-VBI
algorithm has been executed for all groups, a robust design module
is employed to perform joint processing, updating the dominant components
extracted from the SC-VBI outputs of each group.
\end{itemize}

\subsubsection{SC-VBI Algorithm for Each Group}

Compared to existing CS methods, the SC-VBI algorithm in \cite{SC_VBI}
offers significant benefits when dealing with CS problems with dynamic
grids. By employing low-dimension subspace constrained matrix inversion,
it decreases computational complexity and accelerates convergence.
Therefore, the SC-VBI algorithm is a good choice for this problem.
Since the SC-VBI algorithm for each group is identical to that presented
in \cite{SC_VBI}, we briefly review its main idea here. For the $g$-th
group, given the received signal $\boldsymbol{y}_{g}$ and the corresponding
sensing matrix $\mathbf{\Xi}_{g}\left(\boldsymbol{\varOmega}_{g}\right)=\overline{\mathbf{F}}_{g}\mathbf{A}_{g}\left(\boldsymbol{\varOmega}_{g}\right)$,
the goal of the SC-VBI algorithm is to compute the Bayesian posterior
$p\left(\boldsymbol{x}_{g}\mid\boldsymbol{y}_{g}\right)$ and obtain
the maximum likelihood estimation (MLE) of grid parameters $\boldsymbol{\varOmega}_{g}$,
i.e., $\underset{\boldsymbol{\varOmega}_{g}}{\text{argmax}}\ln p\left(\boldsymbol{\varOmega}_{g}\mid\boldsymbol{y}\right)$.
For notation convenience, the index $g$ is omitted in this subsection
since each group is processed independently, i.e, the received signal
and sensing matrix are denoted by $\boldsymbol{y}\in\mathbb{C}^{R\times1}$
and $\mathbf{\Xi}\left(\boldsymbol{\varOmega}\right)\in\mathbb{C}^{R\times L}$,
respectively.

It is very challenging to obtain the exact posterior $p\left(\boldsymbol{x}\mid\boldsymbol{y}\right)$,
since the factor graph of the associated joint probability model contains
loops. Therefore, the SC-VBI algorithm employs an alternating estimation
(AE) framework that iterates between the SC-VBI and grid estimation
(GE) modules until convergence, where the SC-VBI module calculates
the Bayesian posterior $p\left(\boldsymbol{x}_{g}\mid\boldsymbol{y}_{g}\right)$
for given grid parameters $\boldsymbol{\varOmega}_{g}$ output from
the GE module, while the GE module calculates the MLE of $\boldsymbol{\varOmega}_{g}$
for given estimation of $\boldsymbol{x}$ output from the SC-VBI module.
\begin{itemize}
\item \textbf{SC-VBI Module: }
\end{itemize}
For fixed $\boldsymbol{\hat{\varOmega}}$ obtained from the GE module,
the SC-VBI module introduces a three-layer sparse prior model $p\left(\boldsymbol{x},\boldsymbol{\rho},\boldsymbol{s}\right)$
to model the prior distribution of the sparse signal $\boldsymbol{x}$,
and performs Bayesian inference under the SC-VBI framework to estimate
the parameters $\boldsymbol{v}=\left\{ \boldsymbol{x},\boldsymbol{\rho},\boldsymbol{s}\right\} $,
from which it outputs MMSE estimators $\hat{\boldsymbol{x}}$ of $\boldsymbol{x}$
and and the estimated support $\hat{\mathcal{S}}$.

Specifically, a support vector $\boldsymbol{s}\triangleq\left[s_{1},\ldots,s_{L}\right]^{T}\in\left\{ 0,1\right\} ^{L}$
is introduced to indicate whether the $l$-th element $x_{l}$ in
$\boldsymbol{x}$ is active ($s_{l}=1$) or inactive ($s_{l}=0$),
and $\boldsymbol{\rho}=\left[\rho_{1},...,\rho_{L}\right]^{T}$ denote
the precision vector of $\boldsymbol{x}$ (i.e., $1/\rho_{l}$ denotes
the variance of $x_{l}$. Then the joint distribution of $\boldsymbol{x}$,
$\boldsymbol{\rho}$, and $\boldsymbol{s}$, represented as $p\left(\boldsymbol{v}\right)$,
can be expressed as
\begin{equation}
p\left(\boldsymbol{v}\right)=\underbrace{p\left(\boldsymbol{s}\right)}_{\textrm{Support}}\underbrace{p\left(\boldsymbol{\rho}\mid\boldsymbol{s}\right)}_{\textrm{Precision}}\underbrace{p\left(\boldsymbol{x}\mid\boldsymbol{\rho}\right)}_{\textrm{Sparse\ signal}},\label{eq:p(x,rou,s)}
\end{equation}

The prior distribution $p\left(\boldsymbol{s}\right)$ of the support
vector is used to capture the structured sparsity. For instance, to
capture an independent sparse structure, we can set
\begin{equation}
p\left(\boldsymbol{s}\right)=\prod_{l=1}^{L}\left(\lambda_{l}\right)^{s_{l}}\left(1-\lambda_{l}\right)^{1-s_{l}},\label{eq:iid_support}
\end{equation}
where $\lambda_{l}$ represents the sparsity ratio.

The conditional probability $p\left(\boldsymbol{\rho}\mid\boldsymbol{s}\right)$
is given by
\begin{align}
p\left(\boldsymbol{\rho}\mid\boldsymbol{s}\right) & =\prod_{n=1}^{L}\left(\Gamma\left(\rho_{l};a_{l},b_{l}\right)\right)^{s_{l}}\left(\Gamma\left(\rho_{l};\overline{a}_{l},\overline{b}_{l}\right)\right)^{1-s_{l}},\label{eq:ruoconds}
\end{align}
where $\Gamma\left(\rho;a,b\right)$ denotes a Gamma hyper-prior with
shape parameter $a$ and rate parameter $b$. When $s_{l}=1$, the
variance $1/\rho_{l}$ of $x_{l}$ is $\Theta\left(1\right)$, and
thus the shape and rate parameters $a_{l},b_{l}$ should be chosen
such that $\frac{a_{l}}{b_{l}}=\mathbb{E}\left[\rho_{l}\right]=\Theta\left(1\right)$.
On the other hand, when $s_{l}=0$, $x_{l}$ is close to zero, and
thus the shape and rate parameters $\overline{a}_{l},\overline{b}_{l}$
should be chosen to satisfy $\frac{\overline{a}_{l}}{\overline{b}_{l}}=\mathbb{E}\left[\rho_{l}\right]\gg1$.

The conditional probability $p\left(\boldsymbol{x}\mid\boldsymbol{\rho}\right)$
for the sparse signal is assumed to have a product form:
\begin{equation}
p\left(\boldsymbol{x}\mid\boldsymbol{\rho}\right)=\prod_{l=1}^{L}p\left(x_{l}\mid\rho_{l}\right),
\end{equation}
where each $p\left(x_{l}\mid\rho_{l}\right)$ is modeled as a complex
Gaussian prior distribution:
\begin{equation}
p\left(x_{l}\mid\rho_{l}\right)=\mathcal{CN}\left(x_{l};0,\rho_{l}^{-1}\right),\forall l=1,...,L.\label{eq:xcondruo}
\end{equation}

Based on the above three-layer hierarchical prior $p\left(\boldsymbol{v}\right)$
and the complex Gaussian observation $\boldsymbol{y}$, the SC-VBI
module adopts the low-complexity SC-VBI method \cite{SC_VBI} to calculate
the approximate marginal posteriors $p\left(\boldsymbol{v}\mid\boldsymbol{y}\right)$
for fixed grid parameters $\boldsymbol{\hat{\varOmega}}$. In this
approach, the high-dimensional matrix inversion is replaced with a
low-dimensional, subspace-constrained matrix inversion, whose dimension
equals the current sparsity level, and this operation dramatically
reduces computational complexity. Please refer to \cite{SC_VBI} for
further details of the SC-VBI algorithm, the detailed algorithmic
steps are omitted here due to space limitations.
\begin{itemize}
\item \textbf{Grid Estimation (GE) Module:}
\end{itemize}
For fixed $\hat{\boldsymbol{x}}$ and $\hat{\mathcal{S}}$ output
from the SC-VBI module, the GE module refines the dynamic grid $\boldsymbol{\varOmega}$
by directly maximizing the likelihood function based on the gradient
ascent method. To reduce computational complexity, only the sparse
signals and grid parameters $\boldsymbol{\varOmega}_{\hat{\mathcal{S}}}$
corresponding to the estimated support $\hat{\mathcal{S}}$ are retained
when constructing the likelihood function. In this case, the logarithmic
likelihood is expressed as
\begin{align}
\mathcal{L}\left(\boldsymbol{\varOmega}_{\hat{\mathcal{S}}}\right) & \triangleq-\left\Vert \mathbf{\Xi}_{\hat{\mathcal{S}}}\left(\boldsymbol{\varOmega}_{\hat{\mathcal{S}}}\right)\boldsymbol{x}_{\hat{\mathcal{S}}}\right\Vert ^{2}+C,\label{eq:theta_ML}
\end{align}
where $C$ is a constant, and $\mathbf{\Xi}_{\hat{\mathcal{S}}}\left(\boldsymbol{\varOmega}_{\hat{\mathcal{S}}}\right)\in\mathbb{C}^{R\times\left|\hat{\mathcal{S}}\right|}$
is a sub-matrix of $\mathbf{\mathbf{\Xi}\left(\boldsymbol{\varOmega}\right)}$
with the column indices lying in $\hat{\mathcal{S}}$. Then in the
$i\textrm{-th}$ iteration, the grid parameters are updated as
\begin{align}
\boldsymbol{\varOmega}_{\hat{\mathcal{S}}}^{\left(i\right)} & =\boldsymbol{\varOmega}_{\hat{\mathcal{S}}}^{\left(i-1\right)}+\epsilon^{\left(i\right)}\nabla_{\boldsymbol{\varOmega}_{\hat{\mathcal{S}}}}\mathcal{L}\left(\boldsymbol{\varOmega}_{\hat{\mathcal{S}}}\right)\mid_{\boldsymbol{\varOmega}_{\hat{\mathcal{S}}}=\boldsymbol{\varOmega}_{\hat{\mathcal{S}}}^{\left(i-1\right)}},\label{eq:theta_update}
\end{align}
where $\epsilon^{\left(i\right)}$ is the step size determined by
the Armijo rule. The other grid parameters which are not indexed by
$\hat{\mathcal{S}}$ are set to equal to the previous iteration. We
can apply the above gradient update for $B\geq1$ times, where $B$
is chosen to achieve a good trade-off between the per iteration complexity
and convergence speed.

\subsubsection{Robust Design based on Joint Processing}

The group-wise approach has already provided approximate marginal
posteriors $p\left(\boldsymbol{x}_{g}\mid\boldsymbol{y}_{g}\right)$,
the estimated support $\hat{\mathcal{S}}_{g}$, and an approximate
solution for $\hat{\boldsymbol{\varOmega}}_{g}=\underset{\boldsymbol{\varOmega}_{g}}{\text{argmax}}\ln p\left(\boldsymbol{\varOmega}_{g}\mid\boldsymbol{y}\right)$.
However, due to the finite filter length, the designed band-pass filters
may fail to completely suppress the angular components in the neighboring
sub-intervals, introducing errors into the approximate observation
model in (\ref{eq:low_dim_CS_model}).

To compensate for these approximation errors, a robust design based
on joint processing is incorporated. Specially, for given angle grid
parameters $\boldsymbol{\hat{\varOmega}}_{g},g=1,\ldots,G$ obtained
from group-wise processing, we reconstruct the overall sensing matrix
by retaining only the grid points corresponding to the estimated support
$\hat{\mathcal{S}}_{g}$ in each group:
\begin{equation}
\mathbf{\hat{A}}\left(\boldsymbol{\varOmega}\right)=\left[\mathbf{A}_{\hat{\mathcal{S}_{1}}}\left(\hat{\boldsymbol{\varOmega}}_{\hat{\mathcal{S}_{1}}}\right),\ldots,\mathbf{A}_{\hat{\mathcal{S}_{G}}}\left(\hat{\boldsymbol{\varOmega}}_{\hat{\mathcal{S}_{G}}}\right)\right],\label{eq:new_many_part_A}
\end{equation}
where $\mathbf{A}_{\hat{\mathcal{S}_{g}}}\left(\hat{\boldsymbol{\varOmega}}_{\hat{\mathcal{S}_{g}}}\right)\in\mathbb{C}^{N_{r}\times\left|\hat{\mathcal{S}_{g}}\right|}$
is a sub-matrix of $\mathbf{A}_{g}\left(\hat{\boldsymbol{\varOmega}}_{g}\right)$
with the column indices lying in $\hat{\mathcal{S}_{g}}$. Based on
this overall sensing matrix $\mathbf{\hat{A}}\left(\boldsymbol{\varOmega}\right)$,
the received signal $\boldsymbol{y}$ and the analog beam matrix $\mathbf{F}_{a}$,
we formulate a low-dimensional CS problem as follows:
\begin{equation}
\boldsymbol{y}=\mathbf{F}_{a}\mathbf{\hat{A}}\left(\boldsymbol{\varOmega}\right)\hat{\boldsymbol{x}}+\boldsymbol{w},\label{eq:subspaced_CS_in_joint}
\end{equation}
where $\hat{\boldsymbol{x}}\in\mathbb{C}^{S\times1}$ is the refined
angle-domain channel vector with dimension $S=\Sigma_{g}\left|\hat{\mathcal{S}_{g}}\right|$,
corresponding to the pruned grid after support selection. Then we
can again apply the SC-VBI algorithm in \cite{SC_VBI} to solve this
low-dimensional CS problem.

\subsection{Complexity Analysis\label{subsec:Complexity-Analysis}}

We further compare the computational complexity of the proposed GW-SC-VBI
algorithm with the state-of-the-art SC-VBI algorithm \cite{SC_VBI}.
If we directly apply the SC-VBI algorithm to the original CS model
in (\ref{eq:CS_model}), the per-iteration complexities for the SC-VBI
and GE modules are given by $\mathcal{O}\left(N_{RF}L+S^{3}\right)$
and $\mathcal{O}\left(L^{2}S\right)$, respectively \cite{SC_VBI}.

On the other hand, the complexity of the proposed GW-SC-VBI algorithm
consists of the group-wise processing stage and the joint processing
stage. In the group-wise processing stage, we assume that the observation
model for each group is equally partitioned, such that the sensing
matrix dimension for the $g$-th group is $\frac{N_{RF}}{G}\times\frac{L}{G}$,
and the subspace dimension is $\frac{S}{G}$. Consequently, the per-iteration
complexities of the SC-VBI and GE modules for each group becomes $\mathcal{O}\left(\frac{1}{G^{2}}N_{RF}L+\frac{1}{G^{3}}S^{3}\right)$
and $\mathcal{O}\left(\frac{1}{G^{2}}L^{2}S\right)$, respectively.

In the joint processing stage, after the supports from all groups
are merged, the sensing matrix dimension becomes $N_{RF}\times S$,
and $D$ additional iterations are performed for fine-tuning. As a
result, the per-iteration complexity in this stage is $\mathcal{O}\left(N_{RF}S+S^{3}\right)$
for the SC-VBI module, and $\mathcal{O}\left(S^{3}\right)$ for the
grid update module.

To ensure a fair comparison, we assume that the total number of iterations
for the SC-VBI algorithm is $J_{max}$. We then set the total number
of iterations for the GW-SC-VBI algorithm to $J_{max}+D$, where $J_{max}$
iterations are assigned to the group-wise processing and $D$ additional
iterations are used for fine-tuning in the joint processing stage. 

It is important to note that the complexity of GW-SC-VBI within a
single group during the group-wise processing stage is approximately
$\frac{1}{G^{2}}$ of that in the conventional SC-VBI algorithm. In
massive MIMO systems where the sparsity level $S$ is much smaller
than $N_{RF}$ and $L$, the joint processing stage does not introduce
significant additional computational complexity. Therefore, the proposed
GW-SC-VBI algorithm achieves substantially lower complexity than the
conventional SC-VBI algorithm.

\section{Simulations \label{sec:simulations}}

In this section, we provide numerical results to validate the effectiveness
of the proposed group-wise narrow beam design and the corresponding
GW-SC-VBI channel estimation algorithm. We consider a TDD massive
MIMO system with a PC-HBF structure working at a carrier frequency
of $28$ GHz. The BS is equipped with a UPA of $N_{y}=64$ horizontal
antennas and $N_{z}=72$ vertical antennas, leading to a total of
$N_{r}=4608$ antennas. It is noteworthy that with the same number
of antennas, a UPA occupies a much smaller physical size than a ULA,
making near-field effects negligible in our scenarios. Each RF chain
is connected to a vertical subarray consisting of $M=12$ antennas,
and the number of RF chains is given by $N_{RF}=384$. The channel
is generated by QuaDRiGa channel model \cite{quadriga_channel_model},
consisting of one LOS path and $9$ NLOS paths, and the vertical angles
are constrained within prior interval $\Phi$, ranging from $-\frac{1}{6}\pi$
to $0$.

\subsection{Validation of Group-wise Narrow Beam Design}

In this subsection, we compare the proposed group-wise narrow beam
design with several representative baseline beam designs:
\begin{itemize}
\item \textbf{Wide beam: }A straightforward approach where all compression
vectors are set identically as wide beams, effectively forming a band-pass
filter that covers the entire vertical angle range.
\item \textbf{Random beam:} Most common and widely used baseline approach
in the literature, all elements in the compression vectors are randomly
assigned.
\item \textbf{Proposed beam without antenna grouping pattern optimization
(proposed-without-opt):} The group-wise design is applied in the vertical
dimension, while in the horizontal dimension, antennas are grouped
uniformly without further optimization.
\end{itemize}

\subsubsection{Comparison of Vertical AFs under Different Beams}

\begin{figure}
\begin{centering}
\textcolor{blue}{\includegraphics[width=75mm]{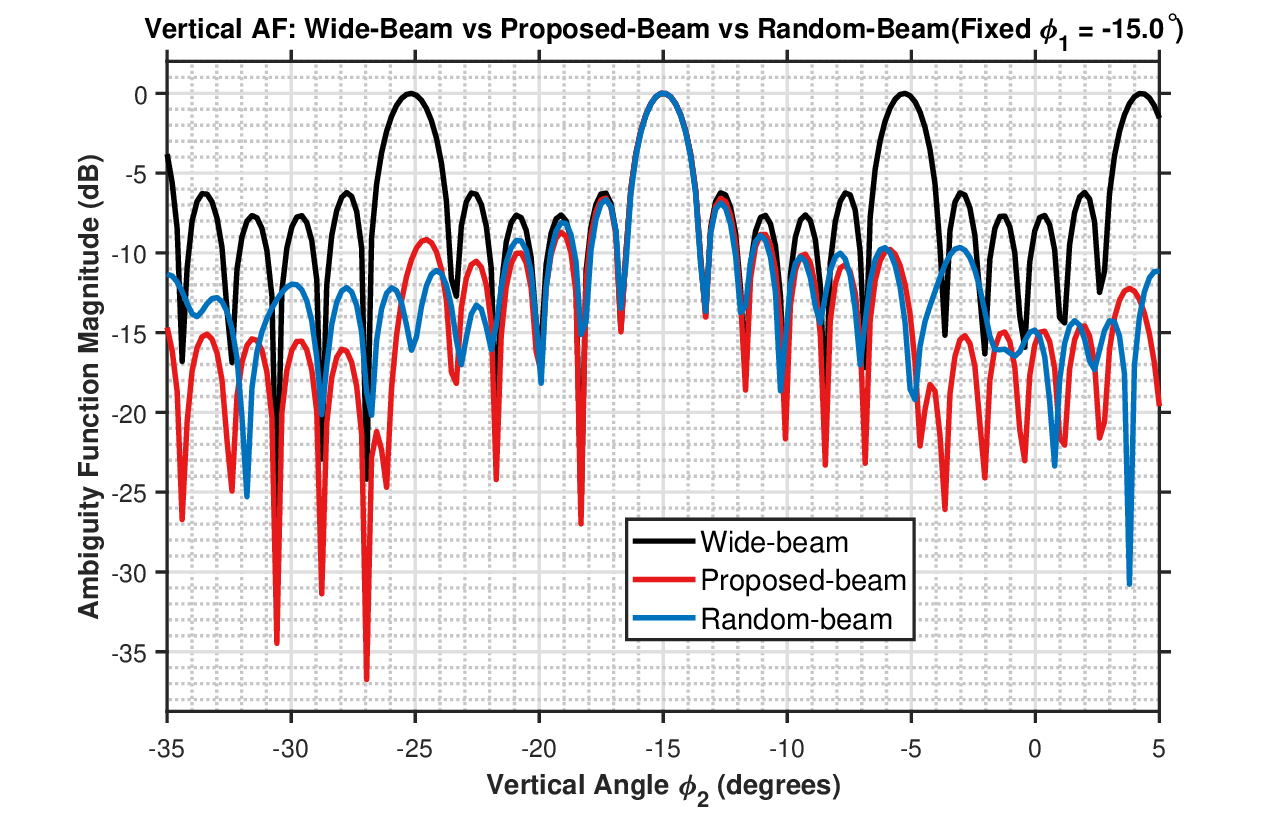}}
\par\end{centering}
\caption{{\small\label{fig:vertical_AF}}Vertical AFs for different analog
beams.}
\end{figure}

\textcolor{black}{Fig. \ref{fig:vertical_AF} illustrates the comparison
of vertical AFs for wide beam, random beam and the proposed group-wise
narrow beam designs. It can be observed that unlike the conventional
wide beam, both the group-wise narrow beam and the random beam exhibit
a much higher main-lobe energy relative to the side lobes. This result
demonstrates that dividing the horizontal antennas into groups and
employing group-wise narrow beams can effectively suppress inter-angle
interference in the vertical domain. }Meanwhile, the random beam cannot
be aligned with the vertical angle range\textcolor{black}{, resulting
in a loss of array gain. Consequently, the proposed design achieves
significantly improved vertical angle estimation performance.}

\subsubsection{Convergence Behavior of EDA in Horizontal Antenna Grouping}

\begin{figure}
\begin{centering}
\textcolor{blue}{\includegraphics[width=75mm]{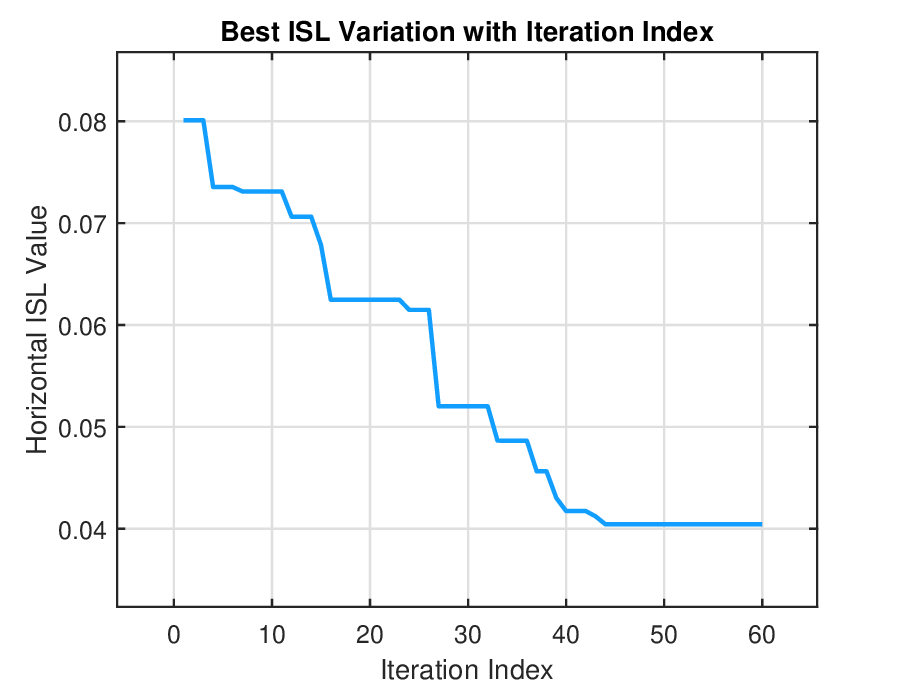}}
\par\end{centering}
\caption{{\small\label{fig:EDA_convergence}}Convergence behavior of the EDA.}
\end{figure}

Fig. \ref{fig:EDA_convergence} illustrates the convergence behavior
of the adopted EDA algorithm for antenna grouping pattern optimization.
As shown, the EDA algorithm converges within $50$ iterations, with
the horizontal ISL value reduced from $0.08$ at initialization to
$0.04$ after convergence. This indicates that the optimized grouping
pattern achieves significantly stronger interference suppression compared
to the random grouping pattern.

\begin{figure}
\begin{centering}
\textcolor{blue}{\includegraphics[width=75mm]{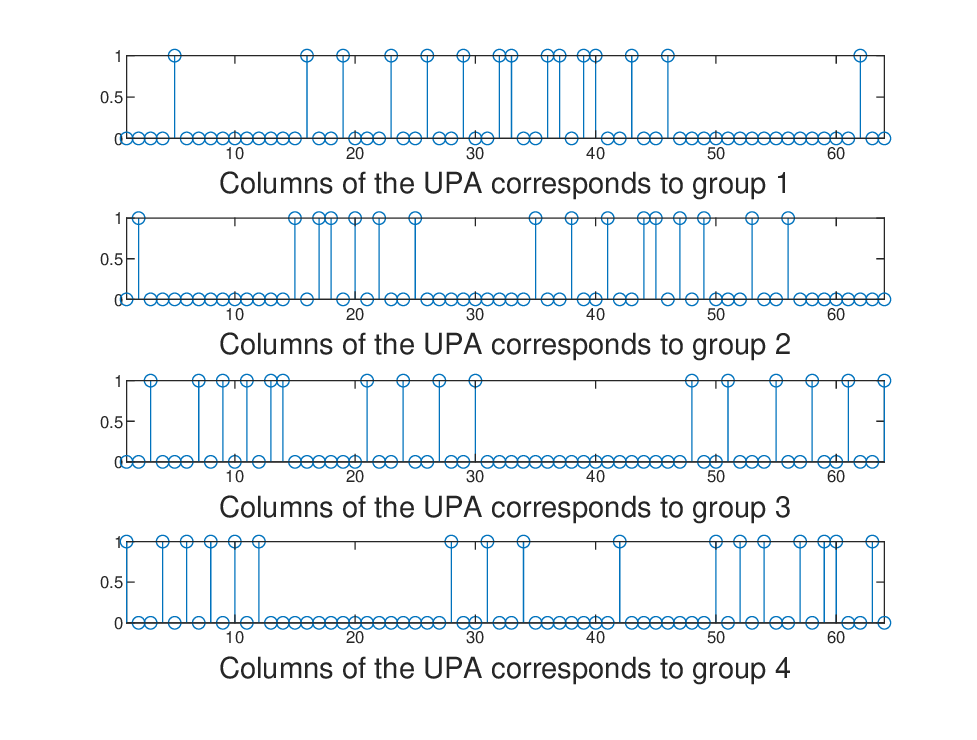}}
\par\end{centering}
\caption{{\small\label{fig:antenna_grouping_pattern}}An illustration of the
optimized antenna grouping pattern.}
\end{figure}

Fig. \ref{fig:antenna_grouping_pattern} further depicts the resulting
grouping pattern when $N_{y}=64$ horizontal antennas are divided
into $G=4$ groups. Unlike the uniform pattern, where antennas in
each group occupy adjacent positions, the optimized pattern spreads
the column indices of each group across the entire array. This configuration
generally corresponds to improved resolution capability, as the effective
aperture of each group is maximized.

\subsubsection{Comparison of Horizontal AFs under Different Antenna Grouping Patterns}

\begin{figure}
\begin{centering}
\textcolor{blue}{\includegraphics[width=75mm]{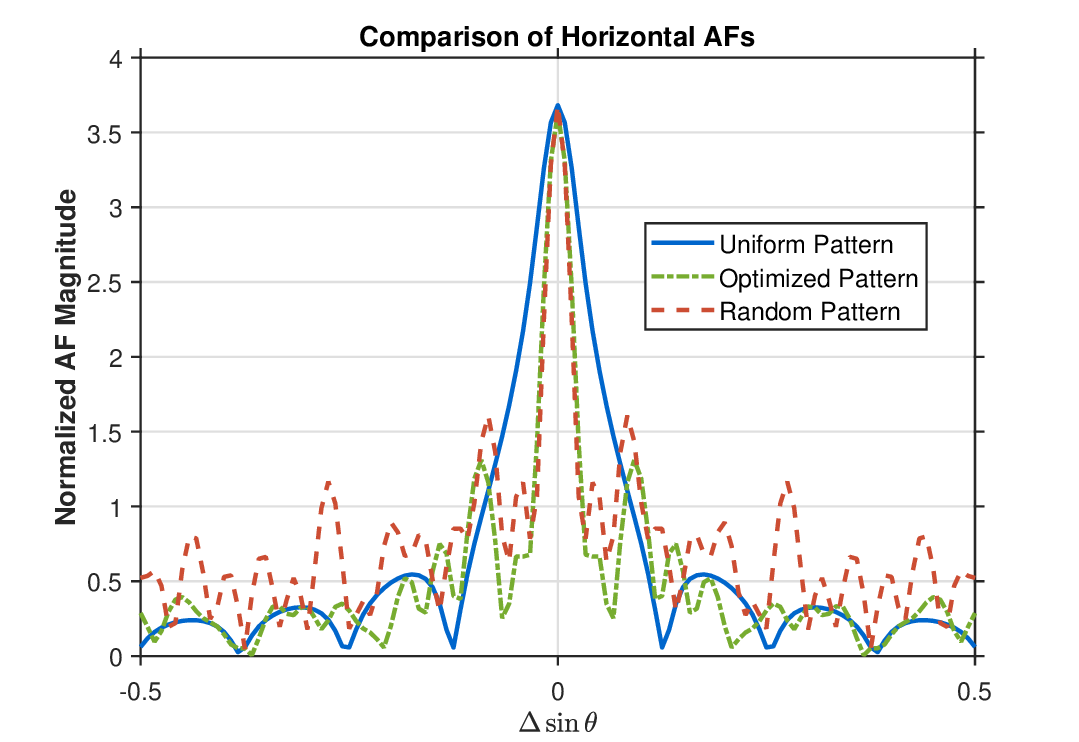}}
\par\end{centering}
\caption{\label{fig:horizontal_AF}Horizontal AFs for three different antenna
grouping patterns.}
\end{figure}

Fig. \ref{fig:horizontal_AF} illustrates the horizontal AFs corresponding
to the uniform, random, and optimized antenna grouping patterns. It
can be observed that both the random pattern and the optimized pattern
yield AFs with significantly narrower main-lobes compared to the uniform
pattern, indicating enhanced resolution performance. This conclusion
is consistent with that shown in Fig. \ref{fig:antenna_grouping_pattern}.
Moreover, the optimized pattern exhibits much lower side-lobe levels
than the random pattern, as seen from the reduced energy in the side-lobes.
This demonstrates improved interference suppression, which agrees
with the results shown in Fig. \ref{fig:EDA_convergence}. In summary,
the proposed optimization achieves a better balance between resolution
and interference suppression, thereby enhancing horizontal angle estimation
accuracy.

\subsubsection{Impact of SNR}

\begin{figure}
\begin{centering}
\textcolor{blue}{\includegraphics[width=75mm]{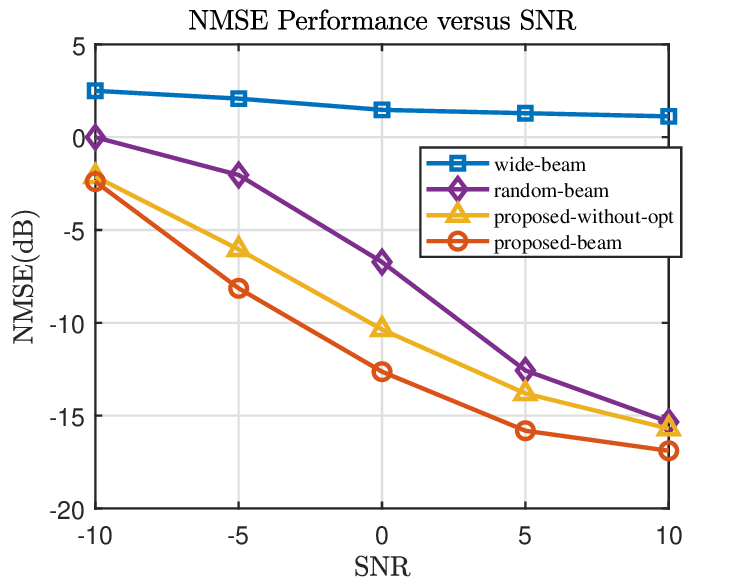}}
\par\end{centering}
\caption{\label{fig:SNR_and_beam}NMSE of channel estimation versus SNR under
different beams.}
\end{figure}

Fig. \ref{fig:SNR_and_beam} shows the NMSE performance of channel
estimation under different analog beam designs across a range of SNR
levels. For fair comparison, all beams are based on the SC-VBI algorithm
for channel estimation. The proposed group-wise narrow beam consistently
outperforms the baseline random beam at all SNR levels. This is because
the group-wise narrow beam is well aligned with the vertical angle
range, thereby preserving array gain. Additionally, optimizing the
antenna grouping pattern in the horizontal dimension provides further
improvement by achieving a superior trade-off between resolution and
interference suppression. As a result, the overall channel estimation
accuracy is significantly enhanced.

\subsubsection{Impact of the Number of Paths}

\begin{figure}
\begin{centering}
\textcolor{blue}{\includegraphics[width=75mm]{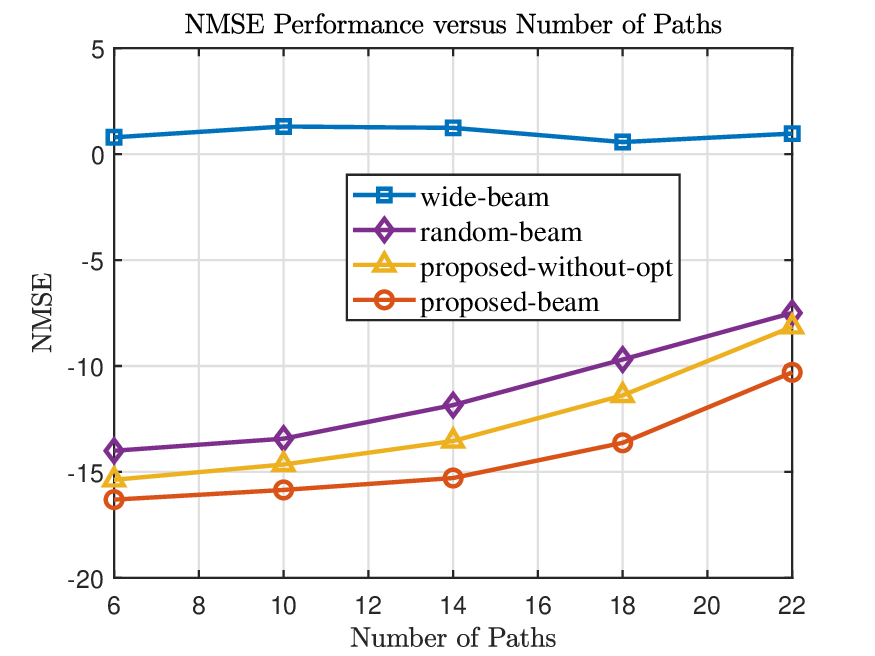}}
\par\end{centering}
\caption{\label{fig:channel_path_and_beam}NMSE of channel estimation versus
the number of paths under different beams.}
\end{figure}

\textcolor{black}{To assess the robustness of the proposed beam design,
the number of paths in the channel is changed from $6$ to $22$,
and SNR is set to $5$dB. All algorithms employ the SC-VBI algorithm
in \cite{SC_VBI} for channel estimation, and simulation results are
summarized in Fig. \ref{fig:channel_path_and_beam}. As expected,
the channel estimation performance of all algorithms degrades as the
number of paths increases. Nevertheless, the proposed group-wise narrow
beam still shows lower NMSE than the baseline random beam across all
path numbers. Moreover, the optimized antenna grouping pattern in
the horizontal dimension continues to provide additional gains. These
phenomenons confirm the effectiveness and robustness of the proposed
beam design.}

\subsection{Validation of the Proposed GW-SC-VBI Algorithm}

In this section, we validate the effectiveness of the proposed GW-SC-VBI
channel estimation algorithm through numerical experiments. The baseline
channel estimation algorithms included for comparison are as follows:
\begin{itemize}
\item \textbf{OMP algorithm \cite{OMP}:} A widely used baseline with low
computational complexity, but its estimation accuracy is typically
limited.
\item \textbf{Turbo-VBI algorithm \cite{LiuAn_CE_Turbo_VBI}:} Achieves
accurate performance and high robustness in Bayesian estimation problems,
but suffers from high computational complexity due to high-dimensional
matrix inversions.
\item \textbf{SC-VBI algorithm \cite{SC_VBI}:} Restricts matrix inversion
to the estimated support, significantly reducing computational complexity
while maintaining comparable estimation performance.
\end{itemize}

\subsubsection{Impact of SNR}

\begin{figure}
\begin{centering}
\textcolor{blue}{\includegraphics[width=75mm]{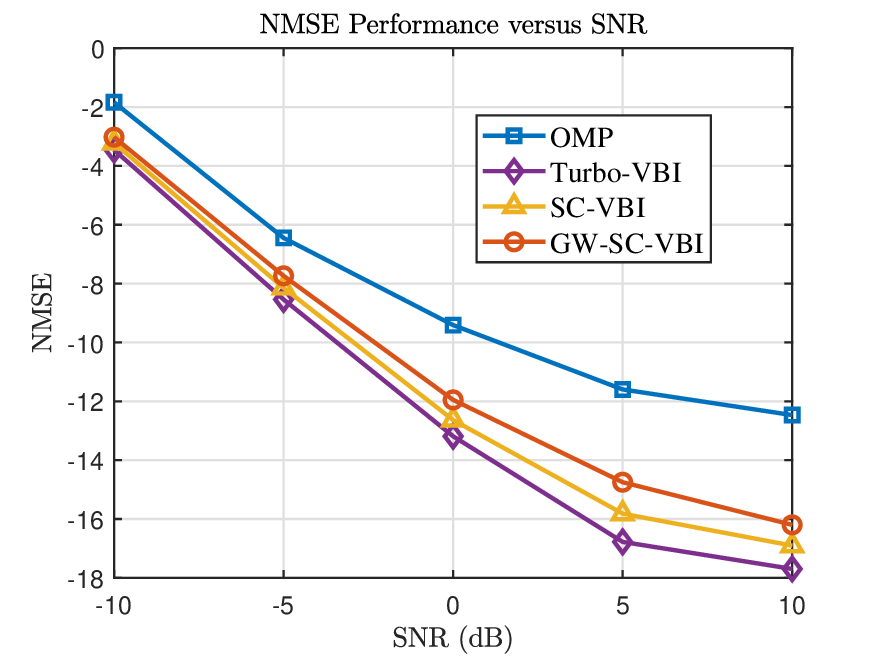}}
\par\end{centering}
\caption{{\small\label{fig:SNR_VBI}NMSE of channel estimation versus the SNR.}}
\end{figure}

Fig. \ref{fig:SNR_VBI} depicts the NMSE performance of the four algorithms
versus SNR, with the same group-wise narrow beam applied to all. The
three VBI-based algorithms notably outperform the baseline OMP algorithm
under all SNR levels. It is important to note that GW-SC-VBI algorithm
achieves comparable estimation performance to Turbo-VBI and SC-VBI,
without significant performance loss but much lower complexity.

Table \ref{tab:CPU_time} measures the CPU time of each algorithm
via MATLAB on a laptop computer with a 2.5 GHz CPU. The results show
that Turbo-VBI incurs the highest computational complexity due to
high-dimensional matrix inversion. Note that the proposed GW-SC-VBI
algorithm reduces the computation time to about one-third of the SC-VBI
algorithm, and only approximately twice that of the simple OMP algorithm,
due to group-wise processing. Given that GW-SC-VBI delivers significantly
better performance than OMP, these results demonstrate that the proposed
GW-SC-VBI algorithm achieves a much better trade-off between accuracy
and complexity.

\begin{table}[t]
\caption{\label{tab:CPU_time}CPU times of different algorithms.}

\centering{}%
\begin{tabular}{|c|c|}
\hline 
\multicolumn{1}{|c|}{Algorithms} & \multicolumn{1}{c|}{CPU times (s)}\tabularnewline
\hline 
OMP & 2.1\tabularnewline
\hline 
Turbo-VBI & 63\tabularnewline
\hline 
SC-VBI & 12.8\tabularnewline
\hline 
GW-SC-VBI & 4.1\tabularnewline
\hline 
\end{tabular}
\end{table}

\subsubsection{Impact of the Number of Paths}

\begin{figure}
\begin{centering}
\textcolor{blue}{\includegraphics[width=75mm]{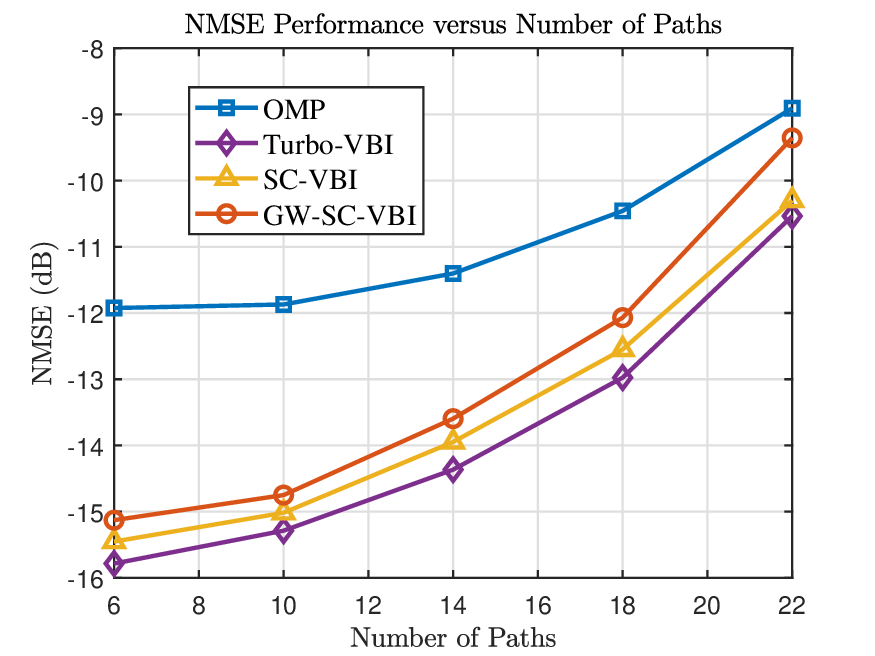}}
\par\end{centering}
\caption{{\small\label{fig:channel_path_VBI}NMSE of channel estimation versus
the SNR.}}
\end{figure}

\textcolor{black}{The channel estimation performance of all algorithms
is further evaluated as the number of channel paths increases from
$6$ to $22$, with SNR fixed at $5$dB and group-wise narrow beam
used. Simulation results summarized in Fig. \ref{fig:channel_path_VBI}
show that estimation accuracy degrades for all algorithms as the number
of paths increases, a trend consistent with larger estimation challenges
in denser multi-path environments. Even so, the GW-SC-VBI algorithm
continues to deliver estimation performance very close to the Turbo-VBI
algorithm and the SC-VBI algorithm, and is consistently superior to
the OMP baseline algorithm across all scenarios. These findings confirm
that the proposed GW-SC-VBI algorithm achieves high estimation accuracy
and strong robustness with substantially reduced computational complexity.}

\section{\textcolor{black}{Conclusion\label{sec:Conclusion}}}

\textcolor{black}{We propose a novel analog beam design and corresponding
channel estimation algorithm for practical HBF massive MIMO systems.
Specifically, the proposed beam adopts a group-wise narrow beam design
in the vertical dimension for effective interference suppression,
and an optimized antenna grouping strategy is introduced in the horizontal
dimension to achieve a better trade-off between resolution and interference
suppression. Based on the proposed group-wise beam, we develop the
GW-SC-VBI algorithm, which effectively reduces computational complexity
by leveraging the group structure to process multiple groups in parallel.
Simulation results demonstrate that the proposed methods achieve robust
and accurate channel estimation performance with much lower complexity
compared to baseline algorithms, confirming their strong potential
for practical application in massive MIMO networks.}

\appendix

\subsection{Derivation of $\mathbf{J}_{\boldsymbol{\mu}_{g}}$ \label{subsec:AppexA}}

We now elaborate on the details of the FIM calculation. For the $g$-th
group, the FIM with respect to $\boldsymbol{\mu}_{g}$ is defined
as:
\begin{equation}
\mathbf{J}_{\boldsymbol{\mu}_{g}}\triangleq\mathbb{E}_{\boldsymbol{y}}\left[-\frac{\partial^{2}\ln\mathit{f}(\boldsymbol{y}|\boldsymbol{\mu}_{g})}{\partial\boldsymbol{\mu}_{g}\partial\boldsymbol{\mu}_{g}^{T}}\right]\in\mathbb{C}^{6\times6},
\end{equation}
where $\boldsymbol{\mu}_{g}\in\mathbb{C}^{6\times1}$ is the parameter
vector defined as $\boldsymbol{\mu}_{g}=\left[\textrm{sin}\left(\theta_{1,g}\right),\textrm{sin}\left(\theta_{2,g}\right),\alpha_{1,g}^{R},\alpha_{2,g}^{R},\alpha_{1,g}^{I},\alpha_{2,g}^{I}\right]^{T}$.
Since the derivations for each entry in $\mathbf{J}_{\boldsymbol{\mu}_{g}}$are
similar, we take the $(1,2)$-th element as an example, given by $\mathbf{J}_{\boldsymbol{\mu}_{g}}\left(1,2\right)=\mathbb{E}_{\boldsymbol{y}}\left[-\frac{\partial^{2}\ln\mathit{f}(\boldsymbol{y}|\boldsymbol{\mu}_{g})}{\partial\textrm{sin}\left(\theta_{1,g}\right)\partial\textrm{sin}\left(\theta_{2,g}\right)}\right]$.

Recall that the complex Gaussian observation model:
\begin{equation}
\boldsymbol{y}=\sum_{g=1}^{G}\left[\mathbf{W}_{g}\otimes\textrm{diag}\left(\boldsymbol{s}_{g}\right)\sum_{k=1}^{2}\alpha_{k,g}\boldsymbol{a}_{R}\left(\theta_{k,g},\varphi_{g}\right)\right]+\boldsymbol{w},
\end{equation}
where $\boldsymbol{w}\sim\mathcal{CN}(0,\sigma^{2}\textbf{I})$. For
brevity, we define $\boldsymbol{b}=\sum_{g=1}^{G}\left[\mathbf{W}_{g}\otimes\textrm{diag}\left(\boldsymbol{s}_{g}\right)\sum_{k=1}^{2}\alpha_{k,g}\boldsymbol{a}_{R}\left(\theta_{k,g},\varphi_{g}\right)\right]$.
According to estimation theory \cite{estimation_theory_book}, for
a complex Gaussian model, the FIM can be calculated as:
\begin{equation}
\mathbf{J}_{\boldsymbol{\mu}_{g}}\left(1,2\right)=\frac{2}{\sigma^{2}}\textrm{Re}\left\{ \frac{\partial\boldsymbol{b}^{H}}{\partial\textrm{sin}\left(\theta_{1,g}\right)}\frac{\partial\boldsymbol{b}}{\partial\textrm{sin}\left(\theta_{2,g}\right)}\right\} ,\label{eq:ori_equation}
\end{equation}
where the first-order derivatives are given by:
\begin{equation}
\begin{array}{c}
\frac{\partial\boldsymbol{b}}{\partial\textrm{sin}\left(\theta_{1,g}\right)}=\mathbf{W}_{g}\otimes\textrm{diag}\left(\boldsymbol{s}_{g}\right)\alpha_{1,g}\frac{\partial\boldsymbol{a}_{R}\left(\theta_{1,g},\varphi_{g}\right)}{\partial\textrm{sin}\left(\theta_{1,g}\right)}\\
\frac{\partial\boldsymbol{b}}{\partial\textrm{sin}\left(\theta_{2,g}\right)}=\mathbf{W}_{g}\otimes\textrm{diag}\left(\boldsymbol{s}_{g}\right)\alpha_{2,g}\frac{\partial\boldsymbol{a}_{R}\left(\theta_{2,g},\varphi_{g}\right)}{\partial\textrm{sin}\left(\theta_{2,g}\right)}
\end{array}.
\end{equation}

The product of the derivative terms can be further simplified using
the approach in (\ref{eq:horizontal_AF}):
\begin{equation}
\begin{aligned}\frac{\partial\boldsymbol{b}^{H}}{\partial\textrm{sin}\left(\theta_{1,g}\right)}\frac{\partial\boldsymbol{b}}{\partial\textrm{sin}\left(\theta_{2,g}\right)}= & M\alpha_{1,g}^{H}\alpha_{2,g}\left[\boldsymbol{s}_{g}^{T}\mathbf{a}_{y}\left(\Delta,\varphi_{g}\right)\odot\left(-\boldsymbol{v}_{\textrm{add}}^{2}\right)\right]\end{aligned}
,\label{eq:cal_deri}
\end{equation}
where $M=\mathbf{a}_{z}\left(\varphi_{g}\right)^{H}\mathbf{W}_{g}^{H}\mathbf{W}_{g}\mathbf{a}_{z}\left(\varphi_{g}\right)$,
and $\boldsymbol{v}_{\textrm{add}}$ is a column vector whose $n$-th
element is $j\pi n\textrm{cos}\left(\varphi_{g}\right)$. 

Substituting (\ref{eq:cal_deri}) into (\ref{eq:ori_equation}), the
final expression of $\mathbf{J}_{\boldsymbol{\mu}_{g}}\left(1,2\right)$
is given by:
\begin{equation}
\begin{aligned} & \mathbf{J}_{\boldsymbol{\mu}_{g}}\left(1,2\right)\\
 & \,\,=K_{\textrm{const}}\sum_{n}n^{2}\textrm{Re}\left\{ \alpha_{1,g}^{H}\alpha_{2,g}\boldsymbol{s}_{g}\left[n\right]\textrm{e}^{j\pi n\left(\textrm{sin}\left(\theta_{2,g}\right)-\textrm{sin}\left(\theta_{1,g}\right)\right)}\right\} 
\end{aligned}
,
\end{equation}
where $K_{\textrm{const}}=\frac{2\pi^{2}M\textrm{cos}^{2}\left(\varphi_{g}\right)}{\sigma^{2}}$
is a constant with respect to $\textrm{sin}\left(\theta_{2,g}\right)$
and $\textrm{sin}\left(\theta_{1,g}\right)$.

The derivations for the other entries of $\mathbf{J}_{\boldsymbol{\mu}_{g}}$
follow similar steps and are omitted here due to space limitations.


\end{document}